\documentclass[fleqn,usenatbib]{mnras}

\usepackage{newtxtext,newtxmath}

\usepackage[T1]{fontenc}

\DeclareRobustCommand{\VAN}[3]{#2}
\let\VANthebibliography\thebibliography
\def\thebibliography{\DeclareRobustCommand{\VAN}[3]{##3}\VANthebibliography}

\usepackage{graphicx}	
\usepackage{amsmath}	

\title[Globular Cluster Kinematics in NGC5846\_UDG1]{Investigating the Ultra-diffuse Galaxy NGC5846\_UDG1 through the Kinematics of its Rich Globular Cluster System}

\author[Haacke et al.]{
Lydia Haacke,$^{1,2}$\thanks{E-mail: lydia.haacke@gmail.com}
Duncan A. Forbes,$^{1,2}$
Jonah S. Gannon,$^{1,2}$
Shany Danieli,$^{3,4}$
Jean P. Brodie,$^{1,2,5}$
\newauthor
Joel Pfeffer,$^{1,2}$
Aaron J. Romanowsky,$^{5,6}$
Pieter van Dokkum,$^{7}$
Steven R. Janssens,$^{1,2,8}$
Maria Luisa Buzzo,$^{1,2,9}$
\newauthor
and Zili Shen$^{7}$
\\
$^{1}$Centre for Astrophysics and Supercomputing, Swinburne University, John Street, Hawthorn VIC 3122, Australia\\
$^{2}$ARC Centre of Excellence for All Sky Astrophysics in 3 Dimensions (ASTRO 3D), Australia\\
$^{3}$Department of Astrophysical Sciences, 4 Ivy Lane, Princeton University, Princeton, NJ 08544, USA\\
$^{4}$School of Physics and Astronomy, Tel Aviv University, Tel Aviv 69978, Israel\\
$^{5}$Department of Astronomy \& Astrophysics, University of California Santa Cruz, CA 95064, USA\\
$^{6}$Department of Physics and Astronomy, San Jos\'e State University, One Washington Square, San Jose, CA 95192, USA\\
$^{7}$Department of Astronomy, Yale University, New Haven, CT 06520, USA\\
$^{8}$Dragonfly Focused Research Organization, 150 Washington Avenue, Santa Fe, 87501, NM, USA
\\
$^{9}$European Southern Observatory, Karl-Schwarzschild-Strasse 2, Garching bei M\"unchen, 85748, Bayern, Germany}

\date{Accepted XXX. Received YYY; in original form ZZZ}

\pubyear{2024}

\begin{document}
\label{firstpage}
\pagerange{\pageref{firstpage}--\pageref{lastpage}}
\maketitle

\begin{abstract}
Recent studies of ultra-diffuse galaxies (UDGs) have shown their globular cluster (GC) systems to be central in unveiling their remarkable properties and halo masses. Deep \textit{HST} imaging revealed 54 GC candidates around the UDG NGC5846\_UDG1 (UDG1), with a remarkable 13 per cent of the stellar light contained in the GC system.
We present a kinematic analysis of UDG1's GC system from observations with the integral field spectrograph KCWI on the Keck II telescope. We measure recessional velocities for 19 GCs, confirming them as members of UDG1, giving a total of 20 confirmed GCs when combined with literature. Approximately 9 per cent of the stellar light are contained just in the confirmed GCs.
We determine the GC system's velocity dispersion to be $\sigma_{\rm GC}$=29.8$^{+6.4}_{-4.9}$ km s$^{-1}$.
We find that $\sigma_{\rm GC}$ increases with increasing magnitude, consistent with predictions for a GC system that evolved under the influence of dynamical friction. The GC system velocity dispersion is constant out to $\sim1R_{\rm eff}$.
Using $\sigma_{\rm GC}$, we calculate $M_{\rm dyn}$=$2.09^{+1.00}_{-0.64}\times$10$^{9}$M$_{\odot}$ as the dynamical mass enclosed within $\sim$2.5 kpc.
The dark matter halo mass suggested by the GC number--halo mass relationship agrees with our dynamical mass estimate, implying a halo more massive than suggested by common stellar mass--halo mass relationships.
UDG1, being GC-rich with a massive halo, fits the picture of a failed galaxy.
\end{abstract}




\section{Introduction}
Large, diffuse, low surface brightness dwarf galaxies have been studied for decades \citep{reaves53, reaves56, impey88, bothun91}. There has been a surge in popularity since their discovery in large numbers in the Coma cluster and the subsequent definition of ultra-diffuse galaxies (UDGs) by \cite{dokkum15}.
By that definition, UDGs have an effective radius $R_{\rm eff}$ larger than 1.5 kpc and central surface brightness $\mu_{g,0}$ fainter than 24 mag arcsec$^{-2}$.
This definition selects some of the most extreme galaxies in terms of size and surface brightness, although other selection criteria have been suggested \citep[for a discussion of selection effects see e.g.][]{vanNest22, li23}.

In seeking to understand their formation, UDGs have been studied through characteristics like the numbers and physical size of their globular cluster (GC) systems \citep{dokkum16, saifollahi21, gannon22, forbes24}, their dynamical mass \citep{dokkum19_df4, danieli19, trujillo19, forbes20, gannon22} and their dark matter (DM) halo profile \citep{dokkum19_df44, forbes24}. They stand out from other dwarf galaxies because many of them have unusually rich GC systems \citep{dokkum16, lim18, forbes20}.
To explain the formation of these extreme objects the `failed galaxy' scenario has been suggested \citep{dokkum15, lim18, danieli22}, for which rich GC systems and massive haloes are expected \citep{forbes20}. In this scenario, star formation in the UDG-to-be is interrupted and it quenches early. This could be caused by early infall into a dense environment and ram pressure stripping of the galaxy's gas component \citep{dokkum15, koda15, yozin15, benavides21}, although it is possible that the galaxy is quenched through other means before falling into a cluster \citep{forbes23}. As a result of the early quenching, a `failed galaxy' has a lower stellar mass than otherwise expected for its halo mass.

The total mass content of a galaxy ($M_{200}$) and its number of GCs ($N_{\rm GC}$) are connected through the total mass contained in the GC system \citep{spitler-forbes09, harris17}. This relationship has been shown to hold for a wide range of halo masses, extending on average into the dwarf galaxy regime.
Assuming an average mass per GC, the GC system mass can be converted to $M_{200}$ via the log-linear, empirical $N_{\rm GC}-M_{200}$ relationship \citep{burkert20}:
\begin{equation}
    M_{200}=5 \times 10^9 M_{\odot} \times N_{\rm GC}.
\end{equation}
A halo mass estimate can also be obtained through the stellar mass-halo mass (SMHM) relation, of which there are multiple variations \citep[e.g.][]{moster13, moster18, behroozi13, behroozi19, danieli23, thornton23}. It is not clear which of these relationships can be applied to UDGs, as the total halo mass based on their number of GCs is not necessarily consistent with the SMHM relation \citep[e.g.][]{beasley16, lim18, forbes20, toloba23, forbes24}.

Apart from the halo mass, the shape of UDGs' DM profile is also not well constrained, specifically whether they have a cusp \citep[e.g.][]{nfw96} or a core \citep[e.g.][]{burkert95, dicintio14, read16, read17}.
The halo profile of the well-studied UDG DF 44 \citep{dokkum16, dokkum17, dokkum19_df44, saifollahi21} is constrained through the measurement of an increasing radial stellar velocity dispersion profile. \cite{dokkum19_df44} found that a cored profile is slightly preferred over a cuspy NFW profile. Their corresponding mass, however, is consistent with both the SMHM and the $N_{\rm GC}$-$M_{200}$ relationship and does not resolve which relationship can be applied to UDGs.

The NGC 5846 group hosts a UDG which is particularly interesting in the context of its GC system, NGC5846\_UDG1 (UDG1). The group itself \citep{mahdavi05, eigenthaler-zeilinger10, marino16} is at a distance of 25 $\pm$ 4 Mpc \citep{mahdavi05}. This close distance allows spectroscopic study of the members' GC systems that is not possible for DF44 and other UDGs in the Coma cluster. UDG1 was first classified as a UDG by \cite{forbes19}. It is extremely GC-rich \citep{mueller21, forbes21, danieli22} and its stellar body and GCs have matching ages, metallicities \citep[hereafter M20]{mueller20} and colours \citep[hereafter D22]{danieli22}. Based on their finding of $\sim$13 per cent of the stellar mass contained in GCs, D22 suggested UDG1 might be a failed galaxy that formed in a short, intense burst of star formation, which was largely confined to its GCs. An intense episode of star formation could have quenched the galaxy and prevented further star formation, while the newly formed GCs then dissolved to form a significant fraction of the currently observable stellar body. Recent discoveries of galaxies at very high redshifts with 50 per cent or more of their mass contained in GCs \citep{mowla24, adamo24} have sparked more interest in this possibility.

UDG1's GC system has been studied with the VLT Survey Telescope \citep{forbes19} and two separate \textit{HST} programmes \citep{mueller21, danieli22}, as well as with MUSE (M20). 
Imaging-based estimates for its GC numbers are available from \cite{mueller21, forbes21, marleau24} and \cite{danieli22}. Until now, only 11 of the GC candidates have been spectroscopically confirmed (M20).

\cite{mueller21} estimated $N_{\rm GC} = 26 \pm 6$ from single-orbit \textit{HST}/ACS observations, with GC candidates based on the colours and sizes of M20's confirmed GCs.
\cite{forbes21} estimated $N_{\rm GC} \sim 45$ from ground based imaging with the VLT Survey Telescope. Their estimate is based on the 20 GC candidates they found in that imaging \citep{forbes19}. They assumed the peak of the GCLF at $M_{V}(\rm TO)=-7.3$ mag, typical for dwarf galaxies \citep{miller-lotz07}, and inferred roughly 45 GCs for the whole GC system at an assumed distance of 24.89 Mpc.
\cite{marleau24} estimated $N_{\rm GC}$ from the same single orbit \textit{HST}/ACS observations as \cite{mueller21}, also basing their candidates on the properties of M20's spectroscopically confirmed GCs. Assuming a distance of 20.3 Mpc and using the GCLF, they arrived at $N_{\rm GC}=38 \pm 7$ GCs.
D22 estimated $N_{\rm GC}$ from two orbits of \textit{HST}/WFC3 observations. The GC candidates were selected in the F606W and F475W filters. They fitted a GCLF to their GC candidates and, assuming a distance of 26.5 Mpc, found $54 \pm 9$ GCs.

All of the imaging estimates imply a rich GC system and, applying the $N_{\rm GC}$-$M_{200}$ relationship, a total halo mass of >10$^{11}$ M$_{\odot}$.
This is an overly massive halo when compared to expectations based on the SMHM \citep{moster18, behroozi19, danieli23}. 
\cite{forbes24} examined the implications of cuspy and cored profiles using UDGs from the literature with more than 20 GCs and a measured velocity dispersion for either the stars or the GC system. With the velocity dispersion they calculated the dynamical masses using the mass estimator from \cite{wolf10} and extrapolated the halo mass from that. For UDGs they found GC-rich UDGs to be dark matter dominated within the half-light radius, and favoured cored profiles to reproduce the high total halo masses predicted through the high GC counts.

A massive DM halo for UDG1 is also supported by the findings of \cite{bar22}, who analysed the mass segregation in the GC system. They found mass segregation arising naturally for different initial distributions of GCs and dynamical friction (DF) as a natural explanation for the observed segregation.
\cite{liang24} modelled UDG1's GC system and also found DF to be an explanation for the present day distribution of UDG1's GCs. For the high GC luminosity fraction observed, their conclusions rely on the assumption that all the stellar mass formed initially in GCs. They predicted inwards migration of GCs under the influence of DF and the GC system velocity dispersion to be lower than the stellar velocity dispersion. The eleven confirmed GCs from M20 have a measured velocity dispersion of $\sigma_{\rm GC,M20}$=9.4$^{+7.0}_{-5.4}$ km s$^{-1}$, lower than, but within the joint uncertainties of, the stellar velocity dispersion $\sigma_{\ast}$=17 $\pm$ 2 km s$^{-1}$ \citep{forbes21}.
So far, there are only three other UDGs with both $\sigma_{\ast}$ and $\sigma_{\rm GC}$ measured \citep{gannon24_cat}, NGC 1052-DF2 \citep{emsellem19, danieli19, lewis20}, NGC 1052-DF4 \citep{dokkum19_df4, shen23} and VCC 1287 \citep{gannon20}. For all, $\sigma_{\ast}$ and $\sigma_{\rm GC}$ lie within the uncertainties of each other.

In this work, we study UDG1 through spectroscopic data obtained with the integral field spectrograph KCWI (Keck Cosmic Web Imager, \citealp{morrissey18}) with the aim of confirming more members of the GC system and analysing the galaxy's dynamics. In Section \ref{sec:data}, we describe the observations, data reduction, how sources were selected and their spectra extracted. In Section \ref{sec:results}, we describe the results of the data analysis, including the GC number, GC system's velocity dispersion, and the galaxy's dynamical mass. In Section \ref{sec:discussion}, we discuss the implications of the number of confirmed GCs for the total $N_{\rm GC}$ and the influence of dynamical friction on the GC system. Section \ref{sec:summary} summarises our results and conclusions.
We base our analysis on the GC candidates from D22. Therefore, we refer to D22 for UDG1's stellar mass ($M_{\ast}\sim 1.2 \times 10^8$ M$_{\odot}$), effective radius ($R_{\rm eff}=1.9$ kpc), distance ($d=26.5$ Mpc), centre coordinates (RA=226.334525$^{\circ}$, Dec=1.81295$^{\circ}$), S\'{e}rsic index ($n=0.61$) and the magnitudes of the GCs throughout this work. We note, however, that slightly different estimates can be found in \cite{mueller21}, most noteably a larger S\'{e}rsic index ($n=0.73$), smaller effective radius ($R_{\rm eff}=1.7$ kpc) and a higher stellar mass ($M_{\ast} \sim$1.7$\times$10$^8$ M$_{\odot}$) at their assumed distance ($d=21$ Mpc).
Throughout this paper we use the AB magnitude system, refer to projected radii as $R$ and three dimensional radii as $r$, and assume $\Lambda$CDM cosmology with $H_0$=70 km s$^{-1}$ Mpc$^{-1}$.

\section{Data and Methods}\label{sec:data}
\begin{figure}
    \centering
    \includegraphics[width=\columnwidth]{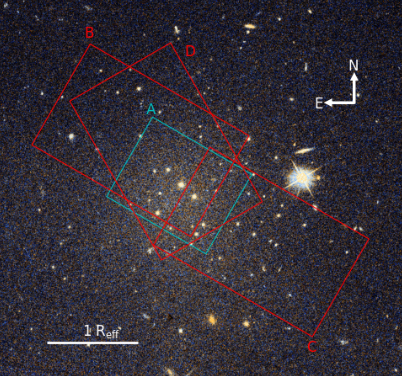}
    \caption{All pointings in which UDG1 was observed with KCWI, overlayed on a colour image from the F475W and F606W filters of the \textit{HST} WFC3/UVIS. Shown in red and blue respectively are the areas covered by observations with the BH3-Large and the BH3-Medium grating of KCWI. The white bar corresponds to one effective radius of the galaxy, $R_{\rm eff}=1.9$ kpc at an assumed distance of 26.5 Mpc \citep{danieli22}. A summary of the configuration and exposure time for each pointing can be found in Table \ref{tab:observations_overview}.}
    \label{fig:pointings}
\end{figure}

UDG1 was observed for a total of $\sim$17 hours with KCWI on the Keck II telescope with the BH3 grating.
Observations were made with both the Medium slicer, which has a field of view of $16.5 \times 20.4$ arcsec$^2$, and the Large slicer, which has a field of view of $33 \times 20.4$ arcsec$^2$. With the BH3 grating the slicers have a spectral resolution of R=9000 (Medium) and R=4500 (Large), respectively.
On 2019, March 30th, May 1st and May 30th, the galaxy was observed for 6 hrs in total with the Medium slicer at the central wavelength of 5110 \AA~and position angle (PA) of 60$^{\circ}$ under programme N061 (PI Romanowsky). The wavelength coverage spans 4861 \AA~ to 5336 \AA~. UDG1 was also observed with the Large slicer at central wavelength of 5080 \AA~and PA=330$^{\circ}$ on 2021, April 15th for 3.3 hours each on two different pointings under programme Y228 (PI van Dokkum) and on 2021, April 16th and 17th with the same central wavelength and PA=120$^{\circ}$ for another 4.5 hours under programme U105 (PI Brodie). The wavelength coverage of these observations spans 4825 \AA~ to 5313 \AA~. Standard star observations were obtained in the respective same configurations.
The different pointings are shown on sky in Figure \ref{fig:pointings} and summarised in Table \ref{tab:observations_overview}. Overall, the observing conditions were clear. However, 40 minutes of exposure have been excluded from the 5.9 hrs of observations with the Medium slicer because of configuration errors and deteriorating weather conditions.

\begin{table}
    \centering
    \begin{tabular}{|c|c|c|c|c|c|}
        \hline
        Pointing & Date & $\lambda_{\rm central}$ & Position & Slicer & Exposure\\
         & & [\AA] & Angle [$^\circ$] &  & Time [h]\\
        \hline
        \hline
        A & 2019/03/30 & 5110 & 60 & Medium & 2.3\\
        A & 2019/05/01 & 5110 & 60 & Medium & 3.0\\
        A & 2019/05/29 & 5110 & 60 & Medium & 0.6\\
        B & 2021/04/15 & 5080 & 330 & Large & 3.3\\
        C & 2021/04/15 & 5080 & 330 & Large & 3.3\\
        D & 2021/04/16 & 5080 & 120 & Large & 1.3\\
        D & 2021/04/17 & 5080 & 120 & Large & 3.1\\
        \hline
    \end{tabular}
    \caption{An overview of all observations of UDG1. All were taken with the BH3 grating of KCWI on the Keck II telescope. From left to right the columns contain the pointing as shown in Figure \ref{fig:pointings}, the date of observation, the central wavelength $\lambda_{\rm central}$, the position angle, the employed slicer and the exposure time.}
    \label{tab:observations_overview}
\end{table}

\subsection{Data Reduction}\label{subsec:data_reduction}
All raw data were processed using the KCWI Python data reduction pipeline\footnote{https://kcwi-drp.readthedocs.io/en/latest/}. The pipeline was set to include barycentric wavelength correction. Automatic sky subtraction and air to vacuum wavelength correction was turned off. Running the pipeline in this configuration results in non-sky subtracted, flux (i.e. standard star) calibrated, barycentrically corrected intensity cubes (henceforth \textquotesingle data cubes\textquotesingle).

Since the world coordinate system from KCWI varied minutely from data cube to data cube, it was corrected to be consistent across all data cubes. For each pointing, a GC was matched in both the data cubes and the \textit{HST} data from D22. Next, a two-dimensional Gaussian distribution was fitted to the flux of the source in each data cube using the fit feature in QFitsView\footnote{https://www.mpe.mpg.de/$\sim$ott/QFitsView/}. The pixel value on which the peak of the distribution was located was then fixed to the coordinates of the GC as determined from the central pixel of the same object in the \textit{HST} imaging.

The data cubes were rebinned from the rectangular spaxels inherent to KCWI to square spaxels (0.29" x 0.29") using the Python package \texttt{MontagePy}\footnote{https://github.com/Caltech-IPAC/Montage/tree/main/python/MontagePy}, which conserves flux. Using the same package, the rebinned data cubes were stacked to result in one combined data cube (henceforth `stacked cube') per slicer. From these stacked cubes, spectra were extracted for each source fulfilling the GC size and colour criteria in D22.

The criteria resulted in spectra for 39 sources, extracted by summing the flux contribution from spaxels within a given aperture around each source. The size of the aperture was chosen to maximise the included flux from the source while minimising included flux from any nearby sources with aperture radii between 0.3'' and 0.5'', corresponding to 1.1 to 1.7 spaxels, with a pixel scale of 0.29''/spaxel. All spaxels were weighted by the fraction of the area included in the aperture. With seeing between $\sim$0.8'' and $\sim$2'' this means that galaxy and sky contributions were minimised. The extraction of spectra with up to 0.65'' (2.3 spaxels) radius was tested, however, it did not noticeably improve the S/N ratio of the spectra and in the case of some fainter GCs, which appear smaller on the data cubes, it worsened the S/N ratio.

Background subtraction was carried out by selecting a nearby region of the same size and approximately the same distance from the galaxy centre. The spectrum from that area was then subtracted from the source spectrum. For a small number of cases in which the resulting spectrum contained a noise spike many hundred times the flux of an average noise fluctuation, the flux value of the affected pixels was replaced with the median flux value.

\subsection{Analysis}\label{sec:data_analysis}
The 39 spectra were fitted with the penalised pixel fitting code, \texttt{ppxf} \citep{cappellari04, cappellari17, cappellari2022}, using the high resolution spectral template library from \cite{coelho14}, as done e.g. in \cite{gannon24, forbes24_pasa}.

The first step was fitting the spectra on a grid of initial guesses for the redshift spaced in $\Delta z=0.000025$ ($\sim$10 km s$^{-1}$) increments from $z=0.006775$ ($\sim$2016 km s$^{-1}$) to $z=0.007625$ ($\sim$2277 km s$^{-1}$). This range is informed by the galaxy recessional velocity having been previously determined to be at $2167 \pm 2$ km s$^{-1}$ \cite[KCWI]{forbes21} and $2156.4 \pm 5.6$ km s$^{-1}$ \cite[MUSE]{mueller20}.

In the allowed velocity range, each initial velocity guess was also run across a grid of additive (\texttt{deg}) and multiplicative (\texttt{mdeg}) degrees in \texttt{ppxf}, ranging from $-1$ (no additive polynomial) to 14 for the former and 1 to 8 for the latter. This method was intended to detect convergence on a common recessional velocity regardless of the initial \textsc{ppxf} input parameters. In addition to requiring the fit to converge on one result over a wide range of input redshifts, we also required it to display at least two absorption lines, usually including either H$\beta$ or the Mg$b$ absorption triplet.
We found in this step that the choice of additive and multiplicative degrees had no influence on the measured recessional velocity.

A measurement of the velocity dispersion of the spectrum is returned by \texttt{ppxf}. Cases in which this dispersion was much smaller than the instrumental resolution, hence realistically not possible to measure, were discarded. Cases in which the exact input redshift was returned without any uncertainty or velocity dispersion were also discarded. The median value of the remaining fits was then used as the initial guess for obtaining the final velocity. The final fits were run with \texttt{ppxf} parameters \texttt{deg}=4 and \texttt{mdeg}=4.

We attempted to determine the uncertainty on the recessional velocities by masking out 2.5 \AA~ at a time across the whole spectrum until each part of the spectrum had been masked once. The mean of these fits is the recessional velocity listed in Table \ref{tab:gc_list}. Determining the standard deviation of all fits as uncertainties yielded values smaller than the uncertainties returned by \texttt{ppxf} itself. Therefore, in Table \ref{tab:gc_list} we chose to instead quote the mean individual uncertainties returned by \texttt{ppxf} from all final fits. We tested these uncertainties on the brightest GC (GC 1), the faintest GC (GC 19), and one of the GCs only measured in the Medium slicer (GC 14). First, we got an empirical estimate of the noise in the spectrum using the Normalised Mean Absolute Deviation ($\sigma_{\rm NMAD}$); then 1000 Monte Carlo realisations of each spectrum were generated and fitted, perturbing the spectrum according to the estimated noise in each realisation. The standard deviation of these 1000 fits was compared to the uncertainty returned by \texttt{ppxf}. For GC 1 and GC 19, the uncertainties returned from this approach are within 1 km s$^{-1}$ of the error returned by \texttt{ppxf}. For GC 14, the error from the $\sigma_{\rm NMAD}$ approach is $\sim$6 km s$^{-1}$ larger, but still of the same order as the \texttt{ppxf} error. The bootstrap test was repeated for all GCs, overall showing agreement with the \texttt{ppxf} uncertainties. Only one GC is not well behaved in these tests, which is marked in Table \ref{tab:gc_list}.

\begin{figure*}
    \centering
    \includegraphics[width=\linewidth]{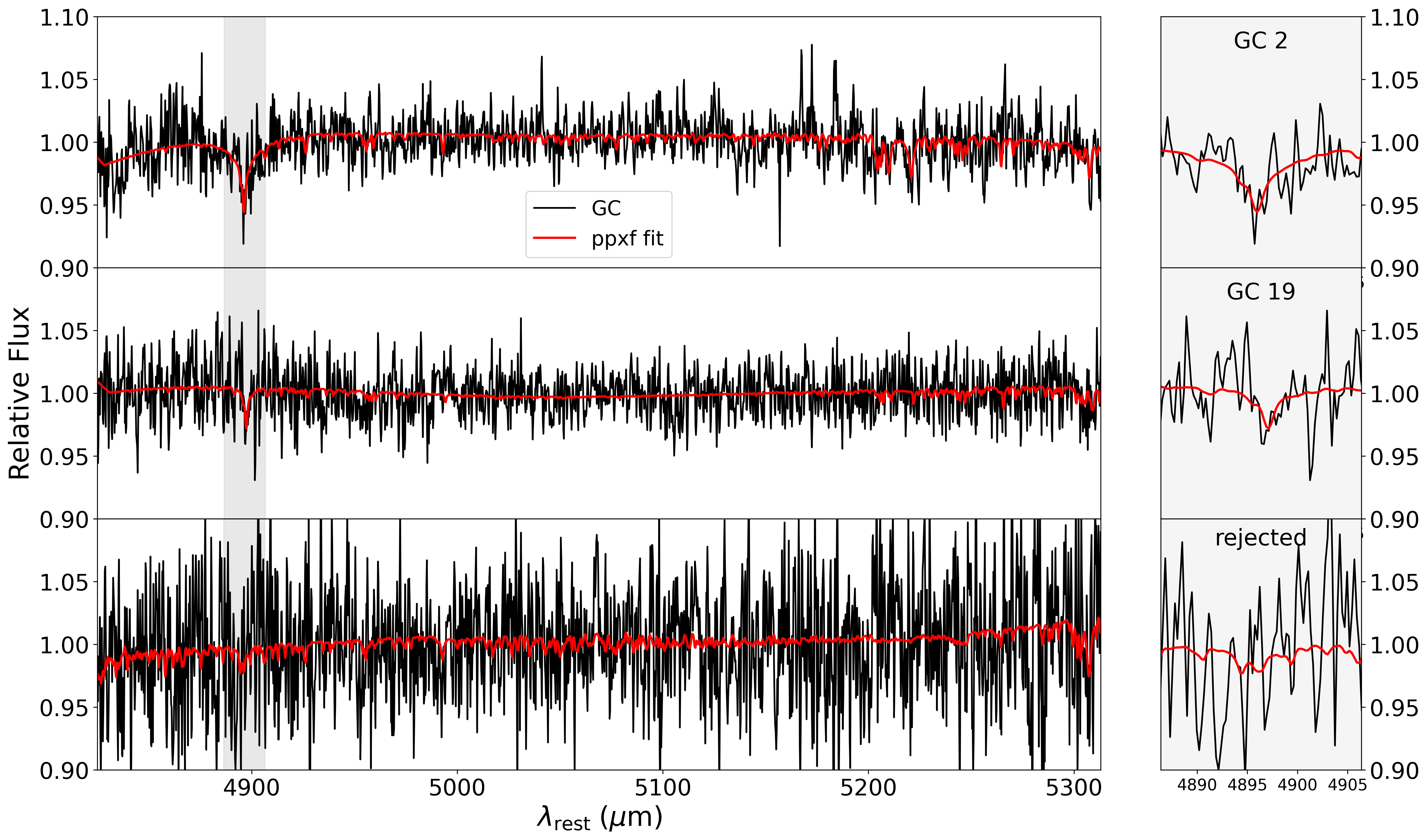}
    \caption{Examples of GC spectra and fits at different S/N ratios. \textit{(Left Column)} The full GC spectrum in black and the \texttt{ppxf} fit in red. \textit{(Right Column)} A zoomed in section around the H$\beta$ line. From top to bottom: an accepted fit of GC 2 at $\rm S/N=8.6$, an accepted fit of GC 19 at $\rm S/N =2.9$, and a rejected fit at $\rm S/N=1.1$.}
    \label{fig:spectra_examples}
\end{figure*}

In total, we recovered recessional velocities for 19 sources within $\pm$100 km s$^{-1}$ of UDG1's recessional velocity, which corresponds to roughly six times the stellar velocity dispersion of 17 $\pm$ 2 km s$^{-1}$ \citep{forbes21}. Examples of accepted and rejected fits can be seen in Figure \ref{fig:spectra_examples}.

For KCWI sources not confirmed as a GC at this stage, we ran an additional fit to check for foreground or background objects. \texttt{Ppxf} does not always return a fit if the initial redshift (recessional velocity) guess is too far from the true redshift, therefore we extended the velocity range from $-200$ km s$^{-1}$ to 5500 km s$^{-1}$ in 40 km s$^{-1}$ increments and repeated the fitting process. None of these fits yielded a result fulfilling all the same criteria applied to the confirmed GCs, hence not confirming any foreground or background objects. We therefore did not find contaminants in the candidates of D22 from our observations (see Figure \ref{fig:pointings} for spatial coverage).

Most of our GCs yield a recessional velocity from the Large slicer. We therefore use this velocity as the final value in Table \ref{tab:gc_list}. Two GCs are only measured in the Medium slicer, and `GC 2' from M20 is not within our area of coverage.
We therefore test potential offsets between the Large slicer and the Medium slicer and between KCWI Large slicer and MUSE velocities. For GCs confirmed in both the Medium and the Large slicer, we find a systematic offset of $10.8 \pm 3.3$ km s$^{-1}$ between the two slicers, with the Medium yielding a systematically higher velocity. Between GCs confirmed in this work and in M20, we find a systematic offset of $7.0 \pm 4.4$ km s$^{-1}$, with MUSE yielding a systematically higher velocity. The statistical significance of the offsets is tested in multiple ways detailed in Appendix \ref{app:bh3m_vs_bh3l} and Appendix \ref{app:kcwi_vs_muse}, respectively. The offsets are applied to the respective GCs throughout this work and in Table \ref{tab:gc_list} the velocities are listed with the offsets applied. Errors on the measured offset are combined in quadrature with the errors of the respective measured recessional velocities.

The galaxy velocity in \cite{forbes21} had been measured with the Medium slicer as well. We applied the offset between the two slicers to it and adopt from here on for UDG1 $v_{\text{\tiny UDG1}}=(2167 - 10.8) \pm 2$ km s$^{-1}=2156.2 \pm 2$ km s$^{-1}$ (noting that after applying the offset it is nearly identical to the value in M20, 2156.4 $\pm$ 5.6 km s$^{-1}$).

\begin{table*}
    \centering
    \begin{tabular}{l|c|c|c|c|c|c|c|c}
        \hline
        ID & RA & Dec & $R/R_{\text{eff}}$ & $v_{\text{los}}$ & S/N & $m_{\rm F606W}$ & $M_{\rm F606W}$ & F475W$-$F606W\\
         & [J2000] & [J2000] & & [km s$^{-1}$] & [\AA$^{-1}$] & [mag] & [mag] & [mag]\\
        \hline
        \hline
        1$\ast$$\dagger$ & 226.3345400 & 1.8129642 & 0.10 & 2137.8 $\pm$ 4.5 & 7.4 & 22.0 & $-10.1$ & 0.38 \\
        2$\ast$$\dagger$ & 226.3339282 & 1.8124165 & 0.10 & 2143.5 $\pm$ 5.5 & 8.6 & 22.5 & $-9.6$ & 0.37 \\
        3$\dagger$ & 226.3365539 & 1.8175039 & 1.23 & 2130.0 $\pm$ 5.6 & 2.5 & 22.8 & $-9.3$ & 0.43 \\
        4$\ast$$\dagger$ & 226.3356742 & 1.8116267 & 0.44 & 2168.8 $\pm$ 5.9 & 9.4 & 22.9 & $-9.2$ & 0.35\\
        5$\dagger$ & 226.3351644 & 1.8136775 & 0.31 & 2147.5 $\pm$ 6.0 & 8.4 & 23.0 & $-9.1$ & 0.41\\
        6$\dagger$ & 226.3335077 & 1.8110790 & 0.42 & 2156.0 $\pm$ 9.4 & 4.8 & 23.1 & $-9.0$ & 0.34 \\
        7$\ast$$\dagger$ & 226.3338263 & 1.8106332 & 0.51 & 2131.3 $\pm$ 6.4 & 8.0 & 23.2 & $-8.9$ & 0.37\\
        GC 2 (M20) & 226.3313573 & 1.8151232 & 0.85 & 2131.5 $\pm$ 23.7 & 4.5 & 23.2 & $-8.9$ & 0.37\\
        8$\ast$$\dagger$ & 226.3364722 & 1.8152582 & 0.79 & 2164.8 $\pm$ 8.7 & 3.7 & 23.4 & $-8.7$ & 0.40\\
        9$\dagger$ & 226.3299469 & 1.8115132 & 1.03 & 2167.0 $\pm$ 7.9 & 2.8 & 23.6 & $-8.5$ & 0.40\\
        10 & 226.3340939 & 1.8101619 & 0.61 & 2176.7 $\pm$ 9.9 & 3.0 & 23.7 & $-8.4$ & 0.38\\
        11$\dagger$ & 226.3358017 & 1.8136011 & 0.43 & 2171.0 $\pm$ 16.2 & 2.9 & 23.8 & $-8.4$ & 0.40\\
        12$\dagger$ & 226.3355595 & 1.8125693 & 0.33 & 2129.5 $\pm$ 7.5 & 2.9 & 23.8 & $-8.3$ & 0.35\\
        13 & 226.3375479 & 1.8173358 & 1.32 & 2142.2 $\pm$ 10.6 & 3.7 & 24.3 & $-7.8$ & 0.43\\
        14 & 226.3350497 & 1.8109007 & 0.48 & 2104.6 $\pm$ 7.5 & 2.6 & 24.5 & $-7.7$ & 0.34\\
        15$^x$ & 226.3328888 & 1.8126786 & 0.30 & 2201.8 $\pm$ 14.1 & 3.0 & 24.5 & $-7.6$ & 0.41\\
        16 & 226.3383585 & 1.8183522 & 1.63 & 2178.7 $\pm$ 13.8 & 2.6 & 24.7 & $-7.4$ & 0.39\\
        17 & 226.3336619 & 1.8157161 & 0.70 & 2180.4 $\pm$ 9.9 & 2.3 & 24.8 & $-7.3$ & 0.43\\
        18 & 226.3340302 & 1.8110026 & 0.41 & 2104.5 $\pm$ 6.2 & 7.8 & 25.2 & $-6.9$ & 0.55\\
        19 & 226.3348586 & 1.8100090 & 0.66 & 2221.5 $\pm$ 9.5 & 2.9 & 25.8 & $-6.3$ & 0.20\\
        \hline
    \end{tabular}
    \caption{Coordinates and measurements for GCs from this work. From left to right, the columns are the internal ID of the GC, the right ascension, the declination, the distance from the galactic centre in units of the effective radius ($R_{\rm eff}=1.9$ kpc, D22), our measured recessional velocity, the signal-to-noise ratio, the apparent magnitude ($m_{\rm F606W}$, D22), the absolute magnitude ($M_{\rm F606W}$) computed assuming a distance of 26.5 Mpc, and the colour. GCs marked with $\dagger$ have counterparts in M20. GCs marked with $\ast$ are those which have recessional velocities from observations with both the Medium and the Large slicer, with both velocities listed in Table \protect\ref{tab:bh3m_bh3l}. Not included in our internal IDs is GC 2 from M20. The value for $v_{\rm rec}$ for that GC is as measured in M20 and corrected by the offset measured in Appendix \ref{app:kcwi_vs_muse}, otherwise all values for the GC are from D22 to stay consistent with the other GCs. GC 15, marked with an $x$, is not well behaved in our tests of the uncertainties described in Section \ref{sec:data_analysis}.}
    \label{tab:gc_list}
\end{table*}

\section{Results}\label{sec:results}
We measured recessional velocities for 19 sources, confirming them as GCs of UDG1. These 19 and one GC from M20 (outside of our area of coverage) are listed in Table \ref{tab:gc_list} along with the internal ID used throughout this work, the position, the distance from the galaxy's centre, the S/N ratio, the apparent magnitude from D22, absolute magnitude assuming the distance of 26.5 Mpc, and the colour.

\begin{figure}
    \centering
    \includegraphics[width=\columnwidth]{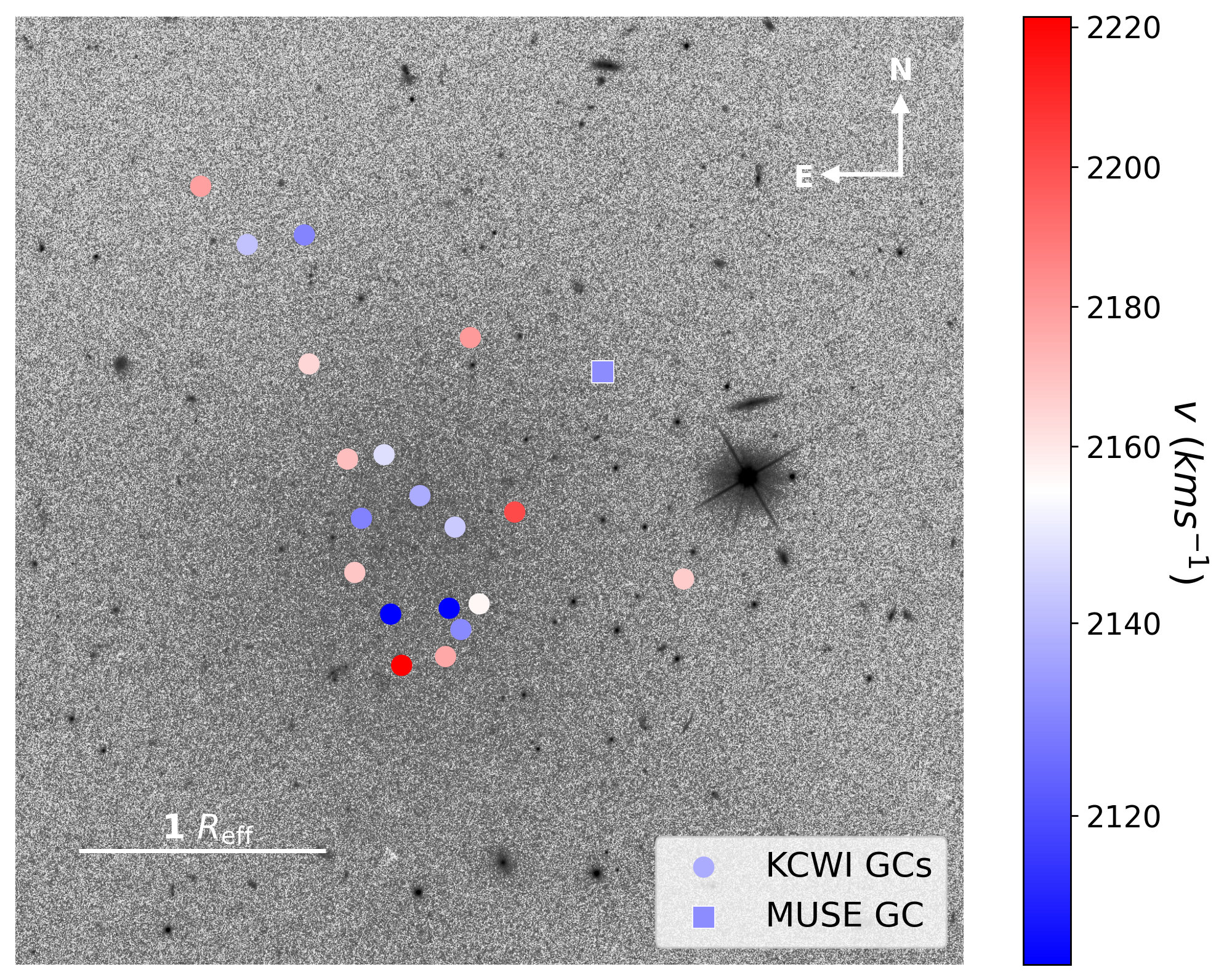}
    \caption{GCs displayed on the F475W and F606W combined image of UDG1, colour coded by recessional velocity and centred on UDG1's centre. Circular points show KCWI GCs from this work, the square point is `GC2' from M20, which is outside the KCWI spatial coverage. The colourmap is centred on the GC system's mean velocity, $\bar{v}_{\rm GC}=2153.9^{+7.1}_{-7.0}$ km s$^{-1}$, which is represented by a white colour. Red colours correspond to GCs redshifted with respect to $\bar{v}_{\rm GC}$, blue colours correspond to GCs blueshifted with respect to $\bar{v}_{\rm GC}$. There is no sign of rotation in the GC system of UDG1.}
    \label{fig:v_gcs_2d}
\end{figure}
Figure \ref{fig:v_gcs_2d} shows the 20 confirmed GCs on sky, colour coded by their recessional velocity. Red sources are redshifted with respect to the GC system's mean velocity and blue sources are blueshifted. `GC 2' from M20 is included in the confirmed GC system, but outside of our area of coverage. There is no visual sign of rotation in the GC system, agreeing with the analysis in M20 for the GC system and in \cite{forbes21} for the stellar body. As done in M20, we run a test for sinusoidal rotation following \cite{lewis20} and find a clear preference for an amplitude of the rotation of 0 km s$^{-1}$. The best fitting velocity dispersion is $\sigma_{\rm GC} = 29.6^{+6.2}_{-4.8}$, in agreement with the value we find for the assumption of a fully dispersion supported system (see Section \ref{sec:results_sigma}). Therefore, we consider the system dispersion supported in line with the previous results from literature.

\subsection{Colours of Globular Clusters}
\begin{figure}
    \centering
    \includegraphics[width=\columnwidth]{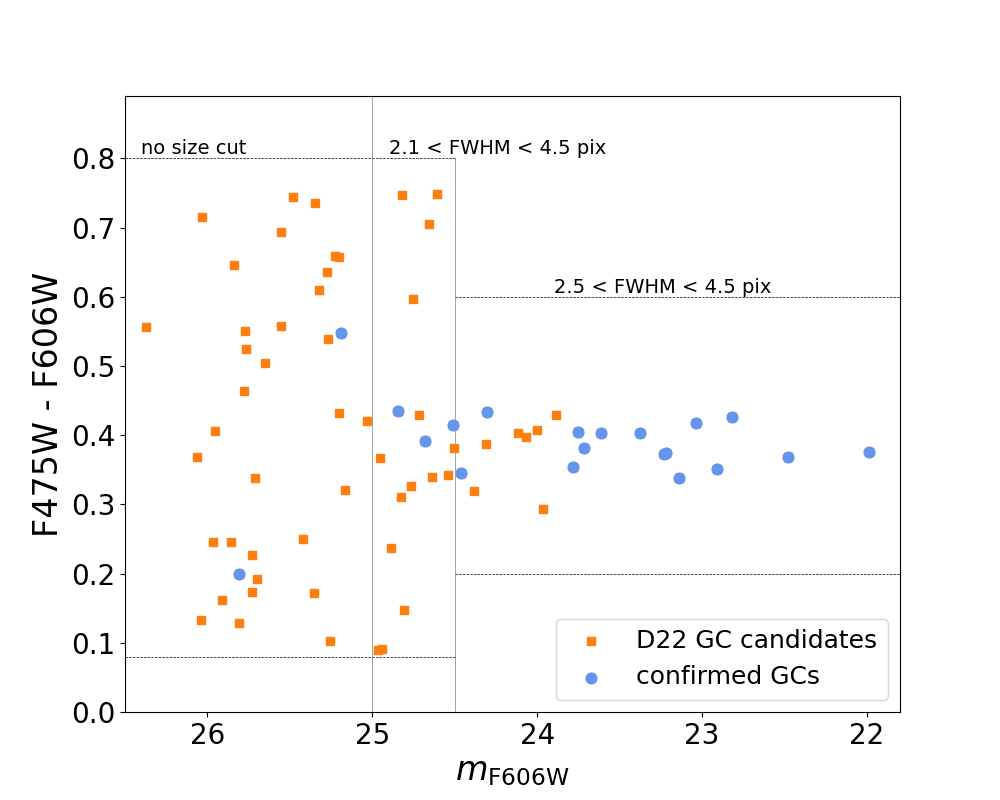}
    \caption{The colour--magnitude diagram of confirmed GCs (blue circles) and candidates fulfilling D22's colour and size criteria (orange squares), following D22's figure 2. The horizontal black dashed lines show the colour limits applied in D22. The vertical black lines separate magnitude ranges within which different FWHM criteria were applied by D22. The allowed FWHM range is labelled in the respective magnitude ranges. Down to $m_{\rm F606W} \sim 23.8$ mag, all GC candidates are confirmed. The confirmed GCs occupy a very tight range of colours, although the spread increases for $m_{\rm F606W} \gtrsim 25$ mag, where the allowed range expands as well in order to account for less precise photometry.}
    \label{fig:gcs_cmd}
\end{figure}
We examined the colours of GC candidates and confirmed GCs. Based on figure 2 in D22, Figure \ref{fig:gcs_cmd} shows the colours of the confirmed GCs, as well as all sources within 2 $R_{\rm eff}$ which fulfill D22's criteria for GC candidates. Sources with $m_{\rm F606W}$ < 24.5 mag have to be in the colour range 0.2 < F475W $-$ F606W < 0.6 and have a full-width half maximum (FWHM) of 2.5 pix < FWHM < 4.5 pix, i.e. be partially resolved at the assumed distance of 26.5 Mpc. Sources with 24.5 mag < $m_{\rm F606W}$ < 25 mag have to be in the widened colour range 0.08 < F475W $-$ F606W < 0.8 but have 2.1 pix < FWHM < 4.5 pix. Sources with $m_{\rm F606W}$ > 25 mag have to be in the colour range 0.08 < F475W $-$ F606W < 0.8 and have no restrictions on the FWHM.

Figure \ref{fig:gcs_cmd} shows the colours and magnitudes of candidates fulfilling the criteria of D22, as well as of the confirmed GCs. Some of the GCs ($m_{\rm F606W} \lesssim 25$ mag) could not be confirmed. This is the case if they are affected by localised noise spikes (as for example the case of one of M20's candidates described in Appendix \ref{app:kcwi_vs_muse}) or outside the area of coverage. The total exposure time also varies across the galaxy, with multiple pointings overlapping in the centre of the galaxy and only individual pointings with less total exposure time available on the outskirts (see also Figure \ref{fig:pointings}).

All of the confirmed GCs occupy a very tight range in colour (0.3 < F475W $-$ F606W < 0.5), with only the faintest two (GC 18 with $m_{\rm F606W}=25.2$ mag and GC 19 with $m_{\rm F606W}$=25.8 mag) outside of that range. Both of the two faint GCs are still fully consistent with the tighter criteria for sources with $m_{\rm F606W}$ < 25 mag. They do, however, spread noticeably further in colour than the brighter GCs.

\subsection{Potential Contamination by Intra-group Globular Clusters}\label{sec:resinterlopers}
\begin{figure}
    \centering
    \includegraphics[width=\columnwidth]{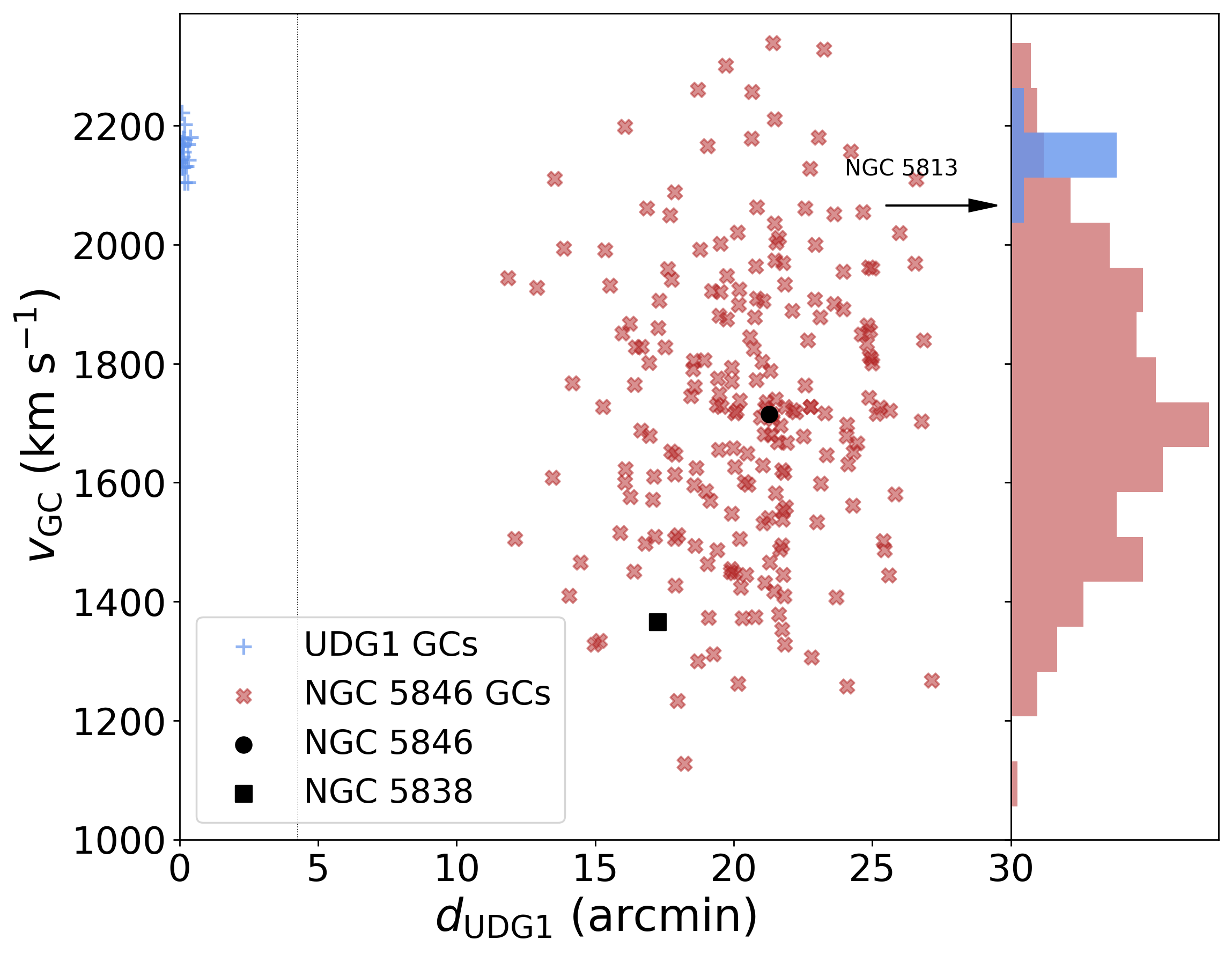}
    \caption{Phase-space diagram for sources of possible interlopers. \textit{(Left)} GCs belonging to NGC 5846 (red crosses) and UDG1 (blue plusses) in velocity space and their projected radial distance from UDG1. Also shown are the position of NGC 5846 (black circle) and NGC 5838 (black square) in phase space. NGC 5813 is marked at its recessional velocity of 2065 km s$^{-1}$ \citep{samsonyan16} with a black arrow but it is projected much further from UDG1 than either NGC 5846 or NGC 5838. The vertical dotted line shows the distance at which the density of NGC 5846's photometric GC candidates falls below 1 GC arcmin$^{-2}$ \citep{zhu16}. \textit{(Right)} The velocity distribution of the GC samples of UDG1 and NGC 5846. They overlap in velocity space, however, they are separated by several arcmin in position. The likelihood of contaminants from NGC 5846, NGC 5813 or NGC 5838 in the GC system of UDG1 is very low.}
    \label{fig:5846_vs_udg1_gcs}
\end{figure}
Potential interlopers in the GC sample in the form of intra-group GCs at the location of UDG1 would most likely be from the dominant giant elliptical galaxies in the group, i.e. from NGC 5846 and NGC 5813. \cite{marleau24} have also noted that the GC system is elongated in the direction of NGC 5838. Figure \ref{fig:5846_vs_udg1_gcs} shows the distance of these three galaxies from UDG1 in phase-space.

NGC 5846 is projected at a distance of $\sim$20 arcmin from the UDG1.
From the SAGES Legacy Unifying Globulars and GalaxieS (SLUGGS) Survey \citep{brodie14} NGC 5846 is known to host over 200 spectroscopically confirmed GCs which are in the velocity range from 900 km s$^{-1}$ to 2400 km s$^{-1}$ and in a similar magnitude range to UDG1's GCs \citep{pota13}. This range includes the recessional velocity of UDG1 and the velocity range of its GCs from this work.
Figure \ref{fig:5846_vs_udg1_gcs} shows the position and recessional velocity of spectroscopically confirmed GCs for NGC 5846 (red) and UDG1 (blue). \cite{zhu16} showed in their figure 1 that the surface number density of GCs falls below 1 GC per arcmin$^2$ at a distance of $\sim$17 arcmin.
At the projected distance from UDG1 ($\sim$20 arcmin), the surface number density of red and blue GCs around NGC 5846 are each already below 0.5 GCs per arcmin$^2$. UDG1's GCs in this work are all contained within an area of 0.85 arcmin$^2$, within which $\sim$0.4 GCs from NGC 5846 would be expected. Less than 10 per cent of the NGC 5846 GCs have velocities higher than 2100 km s$^{-1}$. Combined with the area, this leads to an expected contamination rate of <0.04 GCs at the distance and in the velocity range of UDG1. Therefore, we conclude that there are no likely interlopers from NGC 5846 in our sample despite the overlap in velocity space.

For NGC 5813, the radial velocity of 2065 km s$^{-1}$ \citep{samsonyan16} is similar to UDG1's 2156.2 km s$^{-1}$. \cite{hargis14} estimated the photometric GC density to fall below 1 GC per arcmin$^2$ at a distance of $\sim$14 arcmin. UDG1 is projected at almost five times that distance from NGC 5813 at $\sim$63 arcmin. Hence, we also do not expect interlopers from NGC 5813 despite the similar radial velocities. 

For NGC 5838, there is no existing data on the GC system. The galaxy itself is projected at $\sim 17$ arcmin from UDG1, similar to the distance between NGC 5846 and UDG1. NGC 5838's recessional velocity \citep[$1365 \pm 46$ km s$^{-1}$][]{paturel03} is, however, much lower than UDG1's. It is therefore also not a likely source for interlopers.

Although we cannot completely rule out intragroup GCs in our sample of confirmed GCs, we suggest that it is highly unlikely. 

\subsection{Globular Cluster System Velocity Dispersion}\label{sec:results_sigma}
\begin{figure}
    \centering
    \includegraphics[width=\columnwidth]{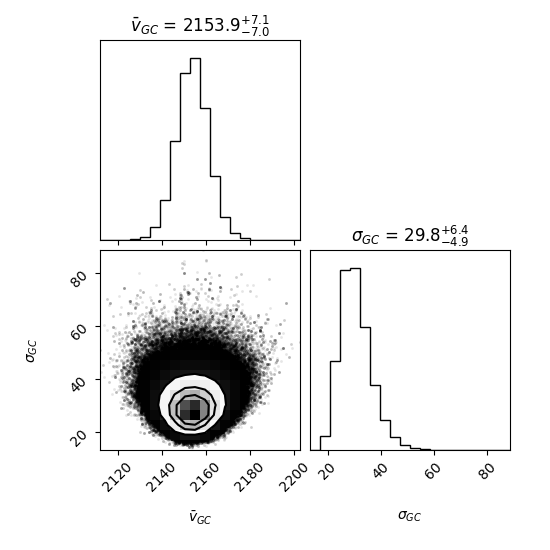}
    \caption{The result of the MCMC fit for the velocity dispersion, $\sigma_{\rm GC}=29.8^{+6.4}_{-4.9}$ km s$^{-1}$, and mean velocity, $\bar{v}_{\rm GC}=2153.9^{+7.1}_{-7.0}$ km s$^{-1}$, of 20 GCs confirmed around UDG1. $\bar{v}_{\rm GC}$ is in good agreement with the value in M20, while $\sigma_{\rm GC}$ is much higher than the corresponding value in M20. $\sigma_{\rm GC}$ is also higher than the stellar velocity dispersion $\sigma_{\ast}$=17 $\pm$ 2 km s$^{-1}$ \citep{forbes21}.}
    \label{fig:kcwi_muse_sigma_corner}
\end{figure}

To determine the velocity dispersion, we ran a Markov Chain Monte Carlo (MCMC) algorithm with Jeffreys prior following \cite{doppel21}. The same method has been applied to other observations in \cite{toloba23}. With the sum of the GC system's velocity dispersion, $\sigma_{\rm GC}$, and the uncertainties on the measurements, $\delta_v$, the average velocity of the GC system, $\bar{v}_{\rm GC}$, and the recessional velocities of the GCs, $v_{\rm obs}$, the log-likelihood function is:
\begin{equation}
    \mathcal{L} =\sum_i \log\Biggl(\frac{1}{\sqrt{2\pi(\sigma_{\rm GC}^2 + \delta^2_{v_i})}}\Biggr) -0.5\frac{(v_{\rm obs,i} - \bar{v}_{\rm GC})^2}{\sigma_{\rm GC}^2 + \delta^2_{v_i}}.
\end{equation}
The chain was implemented using the Python package \texttt{emcee} \citep{foreman-mackey13} with 100 walkers. After a burn-in phase of 1000 steps, it was run for another 20000 steps. The result was determined as the median of the walkers' final positions. The upper and lower uncertainties on $\sigma_{\rm GC}$ are the 84th and 16th percentiles, respectively. We restricted $\sigma_{\rm GC}$ and $\bar{v}_{\rm GC}$, allowing 0 km s$^{-1}$ < $\sigma_{\rm GC}$ < 100 km s$^{-1}$ and $v_{\rm min}$ < $\bar{v}_{\rm GC}$ < $v_{\rm max}$, respectively. $v_{\rm min}$ and $v_{\rm max}$ refer to the lowest and highest recessional velocity measurements of all included GCs.

Figure \ref{fig:kcwi_muse_sigma_corner} shows the result of the MCMC.
We found an average velocity for the GC system of $\bar{v}_{\rm GC}=2153.9^{+7.1}_{-7.0}$ km s$^{-1}$ in very good agreement with M20 ($\bar{v}_{\rm GC}=2150.9^{+5.3}_{-4.9}$ km s$^{-1}$) and a velocity dispersion of $\sigma_{\rm GC}=29.8^{+6.4}_{-4.9}$ km s$^{-1}$, noticeably higher than previously reported by M20 ($\sigma_{\rm GC,M20}=9.4^{+7.0}_{-5.4}$ km s$^{-1}$) and also higher than the stellar velocity dispersion reported by \cite{forbes21}.

In order to examine the possible influence of differences in the method of determining $\sigma_{\rm GC}$, we performed a series of tests on the velocity dispersion. Although the uncertainties were estimated and well-tested as described in Section \ref{subsec:data_reduction}, we perform an additional test here to ensure that a high velocity dispersion is not caused by small uncertainties. We double the uncertainties, leading to a GC system velocity dispersion of $\sigma_{\rm GC}=25.2^{+7.1}_{-5.8}$ km s$^{-1}$, showing that the dispersion remains high and within uncertainties of the value using the quoted uncertainties.

\begin{table}
    \centering
    \begin{tabular}{c|c|c|c|c}
        \hline
        Sample & Source & MCMC & $N_{\rm GC}$ & $\sigma_{\rm GC}$ \\
            &   &   &   & [km s$^{-1}$] \\
        \hline
        \hline
        this work & KCWI & this work & 20 & 29.8$^{+6.4}_{-4.9}$ \\
        M20 & MUSE & M20 & 11 & 9.4$^{+7.0}_{-5.4}$ \\
        M20 & MUSE & this work & 11 & 8.6$^{+7.9}_{-5.8}$ \\
        M20 & KCWI & this work & 11 & 14.0$^{+4.8}_{-3.5}$ \\
        this work, F21 area & KCWI & this work & 12 & 25.9$^{+10.5}_{-6.6}$ \\
        \hline
    \end{tabular}
    \caption{Values for the velocity dispersion $\sigma_{\rm GC}$ for different subsamples. The columns contain, from left to right, the sample, the source, method used to determine $\sigma_{\rm GC}$, the number of GCs in the subsample, and the resulting velocity dispersion. The `source' refers to the instrument used to measure the GCs' velocities. From top to bottom, the samples are: 1) The full sample of confirmed GCs, 2) the result published in M20 3) GCs and velocities from M20 run with our MCMC 4) GCs from M20 run with our MCMC and velocities from KCWI 5) GCs contained in the same area from which \protect\cite{forbes21} determined the stellar velocity dispersion, $\sigma_{\ast}$=17 $\pm$ 2 km s$^{-1}$.}
    \label{tab:velocity_dispersions}
\end{table}

We determined $\sigma_{\rm GC}$ for different subsamples and compared to the result from M20. A summary of the results for all subsamples is in Table \ref{tab:velocity_dispersions}.

We compared our GC system velocity dispersion with the value reported in M20. For this, we first calculated the velocity dispersion with the recessional velocities for the full sample of 11 GCs in M20 using their reported recessional velocities in our dispersion fitting code. We found a systemic velocity of $\bar{v}_{\rm GC}$=2150.7$^{+5.2}_{-4.8}$ km s$^{-1}$ in perfect agreement with M20, and $\sigma_{\rm GC,MUSE}$=8.6$^{+7.9}_{-5.8}$ km s$^{-1}$, slightly lower than reported by M20, $\sigma_{\rm GC,M20}$=9.4$^{+7.0}_{-5.4}$ km s$^{-1}$, but well within the uncertainties. Since M20 used a prior suppressing small values of $\sigma_{\rm GC}$, whereas we used Jeffreys prior, it is expected that our velocity dispersion will be slightly lower.

We then determined the velocity dispersion of the same 11 GCs using instead the velocities measured in this work, with `GC 2' corrected as described in Section \ref{sec:data_analysis}. For this case, we found $\bar{v}_{\rm GC}$ = 2147.7$^{+4.8}_{-4.7}$ km s$^{-1}$ and $\sigma_{\rm GC}$=14.0$^{+4.8}_{-3.5}$ km s$^{-1}$. The systemic velocity stays within the joint uncertainties of the whole system's $\bar{v}_{\rm GC}$. The velocity dispersion is higher than when using M20's velocities, but still remains within the joint uncertainties.

These tests show that there is no strong bias towards lower or higher values for the velocity dispersion due to the method or measured values of the recessional velocities themselves. The choice of prior does, however, influence the outcome.
\cite{doppel21} and \cite{toloba23} examined the difference between a flat prior and Jeffreys prior for simulations and for GCs around Virgo cluster dwarf galaxies, respectively. Both found a flat prior to be biased towards higher velocity dispersions, although this effect becomes negligible for sample sizes of $N_{\rm GC}$ > 10 \citep{doppel21}. In line with their results, we found for UDG1's whole confirmed GC system ($N_{\rm GC}=20$) with a flat prior the same $\sigma_{\rm GC}=29.8^{+6.4}_{-4.9}$ km s$^{-1}$.

\begin{figure}
    \centering
    \includegraphics[width=\columnwidth]{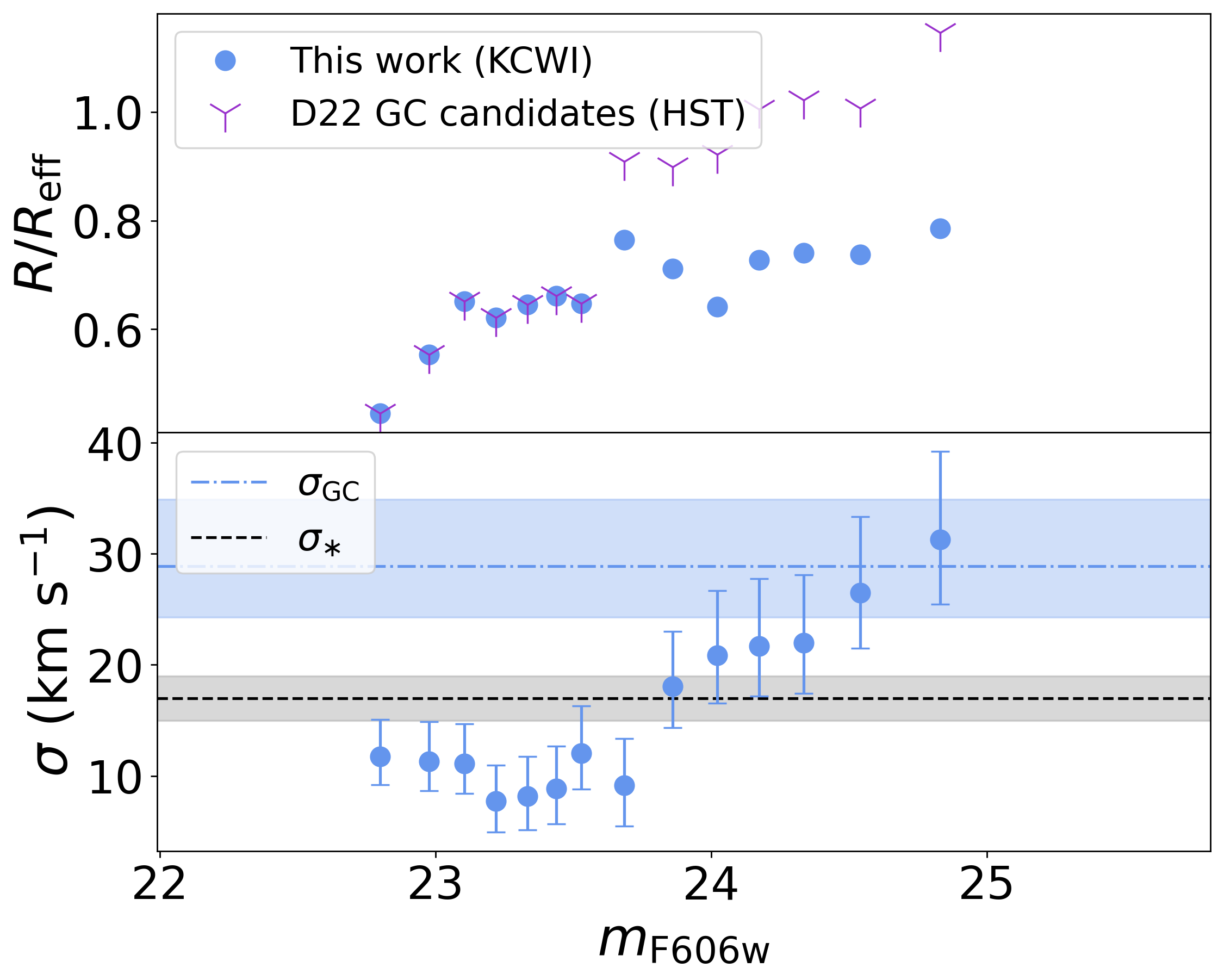}
    \caption{Change in GC system properties with increasing magnitude.
    \textit{(Top)} The mean radius of GCs within each magnitude bin from the spectroscopically confirmed GCs (blue circles) and the imaging candidates from D22 (purple Ys). The mean radius increases, suggesting mass segregation as found in \protect\cite{bar22}. The implications of this segregation are discussed in the context of dynamical friction in Section \ref{sec:df_discussion}.
    \textit{(Bottom)} The velocity dispersion profile of GCs in bins of increasing magnitudes. $\sigma_{\rm GC}$ was determined in bins of seven GCs each, following a `moving window' approach. The black dashed lines shows the stellar velocity dispersion, $\sigma_{\ast}=17\pm2$ km s$^{-1}$, the blue dash-dotted line shows the whole GC system's velocity dispersion, $\sigma_{\rm GC}=29.8^{+6.4}_{-4.9}$ km s$^{-1}$.
    In all panels, the values are plotted against the mean magnitude of all GCs contained in the respective bins. The magnitudes of the GCs range from 22.0 to 25.8 mag. The velocity dispersion increases from $\sigma_{\rm GC}=11.7^{+3.4}_{-2.6}$ km s$^{-1}$ for the brightest seven to $\sigma_{\rm GC}=31.3^{+8.0}_{-5.9}$ km s$^{-1}$ for the faintest seven GCs. The velocity dispersion increasing with fainter magnitudes is consistent with the expected effect of dynamical friction.}
    \label{fig:sig_by_mag}
\end{figure}
Most of the newly confirmed GCs in this work are fainter than the ones confirmed by M20.
We examined the dependence of $\sigma_{\rm GC}$ on the GCs' magnitudes by running the MCMC for bins of multiple sub-samples, sorted by magnitudes. For this, similar to a moving point average, we applied a `moving window' approach. We first determined the velocity dispersion for the brightest seven GCs. This bin size is chosen to minimise sensitivity to individual outliers while also providing enough bins to not miss smaller changes. In the next step, we removed the brightest GC, added the next faintest GC and re-computed the velocity dispersion for those seven GCs. This process was repeated until we reach the faintest seven GCs. The GC system velocity was fixed to the whole system's mean velocity in all bins. The results of this are shown in Figure \ref{fig:sig_by_mag}, where in the bottom panel the velocity dispersion in each bin is shown at the mean magnitude of the included GCs.
The velocity dispersion increases from $11.7^{+3.4}_{-2.6}$ km s$^{-1}$ for the brightest seven GCs to $31.3^{+7.8}_{-5.9}$ km s$^{-1}$ for the faintest seven GCs. These two bins have no GCs in common. The value increases especially with the addition of the faintest five GCs.

To ensure the increase is not caused by the fixed mean velocity, we repeat the same test with the mean velocity as a free parameter. For this case, the mean velocity across the bins remains flat and within the joint uncertainties 

The influence of the GC velocities' uncertainties, $\delta_v$, on the velocity dispersion was tested in three ways.
First, we ran the same moving-window profiles with $\delta_v$ fixed to a constant, artificially small uncertainty for all GCs ($\delta_v$=5 km s$^{-1}$), second with $\delta_v$ fixed to a constant, larger uncertainty for all GCs ($\delta_v$=15 km s$^{-1}$) and third assigning $\delta_v$ proportional to the square of the GCs' magnitudes ($\delta_v = m_{F606W}^2 \times 0.02$). For all cases we found the velocity dispersion profile to be rising with fainter magnitudes. For the tests run with a constant, smaller uncertainty, the velocity dispersion in each bin is overall shifted upwards by $\sim 2$ km s$^{-1}$ and the uncertainties on the dispersion are smaller.
For the tests run with a constant, larger uncertainty, the velocity dispersion of the brightest GCs is overall shifted downwards by $\sim 6$ km s$^{-1}$, whereas the velocity dispersion for the fainter GCs is shifted downwards by $\sim 4$ km s$^{-1}$. The uncertainties on the velocity dispersion decrease in all bins, however, the value is always within the joint uncertainties of the original profile shown in Figure \ref{fig:sig_by_mag}.
For the tests run with the uncertainties proportional to the GCs' magnitude, the velocity dispersion of the brightest GCs is overall shifted downwards by $\sim$ 5 km s$^{-1}$, whereas the velocity dispersion for the fainter GCs does not change noticeably.
The uncertainties on the velocity dispersion increase for the latter two cases so that the shifted velocity dispersion is always within the joint uncertainties of the original profile shown in Figure \ref{fig:sig_by_mag}.
In no case are the last five bins within the joint uncertainties of the first five bins, i.e. the increase of the velocity dispersion with fainter magnitudes is not caused by under- or overestimated uncertainties on the individual GCs' velocities.

\begin{figure}
    \centering
    \includegraphics[width=\columnwidth]{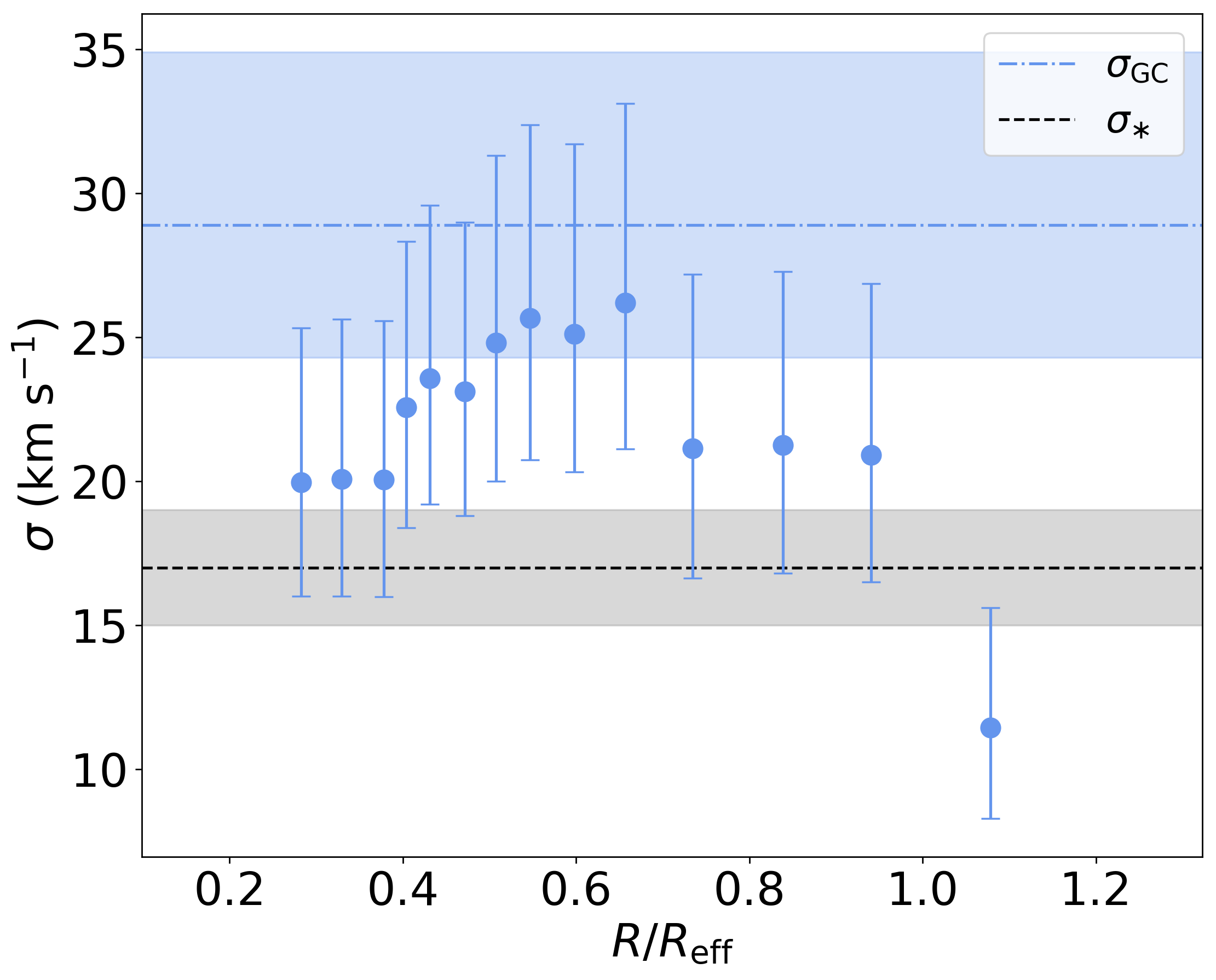}
    \caption{The change in velocity dispersion with increasing projected radius. The dispersion was determined in bins of seven GCs each, following a `moving window' approach. The black dashed lines shows the stellar velocity dispersion, $\sigma_{\ast}$=17$\pm$2 km s$^{-1}$ \citep{forbes21}, the blue dash-dotted line shows the whole GC system's velocity dispersion, $\sigma_{\rm GC}=29.8^{+6.4}_{-4.9}$ km s$^{-1}$.
    The values are plotted against the mean radius of all GCs contained in the respective bins in units of the projected half-light radius, $R_{\rm eff}=1.9$ kpc. The radius of the GCs range from 0.05 $R_{\rm eff}$ to 1.63 $R_{\rm eff}$. The last bin, reaching out to the furthest radius, has a lower velocity dispersion than all other bins due to a GC with a high recessional velocity leaving the 'moving window'. Otherwise, the profile is flat with increasing radius}
    \label{fig:sig_by_r}
\end{figure}

The process was repeated for a radial profile, shown in Figure \ref{fig:sig_by_r}. The GCs are binned by radius in units of $R/R_{\rm eff}$ instead of by magnitude. Again, seven GCs were contained in each bin for which the velocity dispersion was determined, and the dispersion is shown at the mean radius of the GCs contained in each bin.
The change in the velocity dispersion with radius is less linear than with magnitude. There is an initial increase until $\sim0.7 R_{\rm eff}$, although the profile remains consistent with a flat trend within the uncertainties. Notable, the inclusion of the GC most distant from UDG1's centre decreases the dispersion by nearly 10 km s$^{-1}$. We tested the profile shape with independent bins (GC numbers as listed in Table \ref{tab:gc_list}) of GC 1 to GC 7, GC 2 (M20) to GC 13 and GC 14 to GC 19. The shape of the radial GC velocity dispersion profile remains the same in these bins, with the highest value in the second bin and the lowest value in the last bin.
The velocity dispersion profile remains flat with radius, within the uncertainties of the whole system's dispersion and within the uncertainties of the stellar velocity dispersion. The only bin noticeably lower than the overall profile is the last bin (the same is the case for the test with independent bins.) This is due to the GC with the highest recessional velocity leaving the `moving window' for this bin and we do not consider this representative of a true decreasing trend of the velocity dispersion with radius, although a decrease with larger radii could be possible within the errors of the profile.
The radial profile could also be affected by selection and projection effects. Faint GCs are picked up in the central region only when the total exposure time of multiple pointings can be combined. Figure \ref{fig:pointings} also shows sparser sampling of GCs of all magnitudes in the outer regions of the galaxy. The analysis was also done entirely with projected radii, and therefore does not necessarily reflect the true change of the velocity dispersion with the 3D radius.

As with the velocity dispersion profile by magnitude, we tested the influence of $\delta_v$ on the radial velocity dispersion profile. Similarly to before, the uncertainties on the velocity dispersion decrease for a constant, smaller error and increase for a constant, larger error and for an error proportional to the GCs' magnitude. The profile overall shifts upwards by $\sim 1$ km s$^{-1}$ for the constant, smaller $\delta_v$, downwards overall by $\sim 5$ km s$^{-1}$ for the constant, larger $\delta_v$ and downwards overall by $\sim 3$ km s$^{-1}$ for $\delta_v$ proportional to the GCs' magnitudes.
In all cases, the velocity dispersion is within the joint uncertainties of the original profile shown in Figure \ref{fig:sig_by_r}.

Lastly, we tested the sensitivity of the whole confirmed sample's velocity dispersion to outliers. A jackknife procedure was applied where we removed one GC at a time and re-computed $\sigma_{\rm GC}$. The mean of all subsamples determined this way is $\sigma_{\rm GC}$ is 28.9 km s$^{-1}$ with a standard deviation of 1.3 km s$^{-1}$. The largest deviation reduced $\sigma_{\rm GC}$ by $\sim$15 per cent to $24.5^{+5.5}_{-4.2}$ km s$^{-1}$. Within the joint uncertainties, that value still does not overlap with the stellar velocity dispersion. A low number of GCs with velocities far from the system's mean can evidently have a noticeable influence on the result, but removing them does not change our findings qualitatively. All further analysis based on $\sigma_{\rm GC}$ makes use of the whole system's velocity dispersion with Jeffreys prior, $\sigma_{\rm GC}=29.8^{+6.4}_{-4.9}$ km s$^{-1}$.
 
\subsection{Host Galaxy Mass}\label{sec:result_mass}
\begin{figure}
    \centering
    \includegraphics[width=\columnwidth]{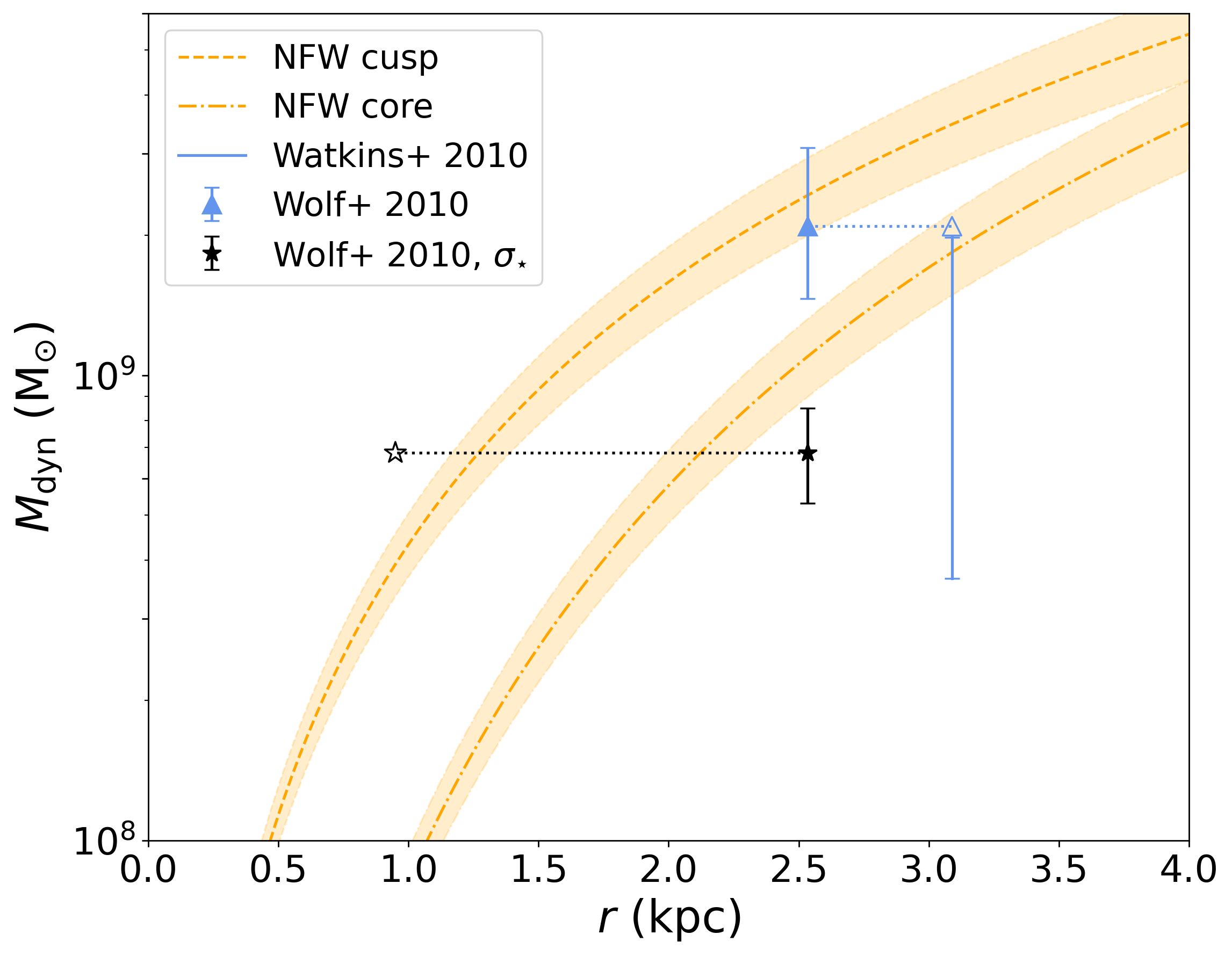}
    \caption{The dynamical masses calculated according to different methods: \protect\cite{wolf10} based on the GC velocity dispersion shown as a blue triangle, based on the stellar velocity dispersion shown as a black star. Empty symbols represent the radius at which the data for each of the estimates actually probes. The estimate according to \protect\cite{watkins10} is shown as a blue capped line, representing a range from the assumption of a cuspy profile (high end) to the assumption of a cored profile (low end). The \protect\cite{watkins10} method is dependent on additional assumptions about additional parameters described in Section \ref{sec:result_mass}, which the \protect\cite{wolf10} is less sensitive to. The estimate based on $\sigma_{\rm GC}$ is higher than the one based on $\sigma_{\ast}$, possibly due to an increasing stellar velocity dispersion with radius.
    Also shown are a cuspy (dashed orange line) and a cored (dash-dotted orange line) halo profile, both with a total mass of $M_{200}=2.7 \times 10^{11}$ M$_{\odot}$ from the $N_{\rm GC}-M_{200}$ relationship. This mass is calculated under the assumption that there are 54 GCs (D22).}
    \label{fig:mdyn_total}
\end{figure}

Figure \ref{fig:mdyn_total} shows dynamical mass estimates with the \cite{wolf10} method with the estimate based on the GC velocity dispersion shown as a blue triangle, the estimate based on the stellar velocity dispersion shown as a black star, and the dynamical mass estimate made with the \cite{watkins10} method as a blue capped line. It should be noted that the \cite{wolf10} method returns the dynamical mass at the de-projected effective radius under the assumption of a representative, flat velocity dispersion with radius. The stellar and GC system velocity dispersion, however, in reality probe the dispersion out to $\sim0.5 R_{\rm eff}$ ($\sim1$ kpc) and $\sim1.6R_{\rm eff}$ ($\sim3.1$ kpc), respectively. These radii are shown as empty symbols in Figure \ref{fig:mdyn_total}.
Also shown are a cuspy (orange dashed) and a cored (orange dot-dashed) DM halo profile, each with a total mass of $M_{200}=2.7^{+2.7}_{-1.4}\times 10^{11}$ M$_{\odot}$, the halo mass implied by the $N_{\rm GC}-M_{200}$ relationship and the 54 GC candidates from D22. The inherent scatter of 0.3 dex \citep{burkert20} dominates the uncertainty in the GC number here and is used as the uncertainty on the mass of UDG1 implied by the $N_{\rm GC}-M_{200}$ relationship throughout this work. The masses and profile were created as described below.

\cite{watkins10} provided equations to estimate the enclosed dynamical mass for a set of discrete tracers, in our case GCs, from their line-of-sight velocities and projected radial distance from the galaxy centre. We used their equation (26) for the scenario in which only projected radii, $R$, and line of sight velocity, $v_{\rm los}$, are available for the mass tracers:
\begin{equation}\label{eq:m_watkins}
    M(r < R)=\frac{C}{G} \langle v^2_{\rm los}R^{\alpha} \rangle.
\end{equation}
$G$ is the gravitational constant and $C$ is a constant depending on the parameters $\alpha$, $\beta$ and $\gamma$, describing the logarithmic slope of the gravitational potential, the orbits of the tracers and the logarithmic slope of the tracer radial density profile, respectively.

For $\gamma$, we followed \cite{beasley16} and \cite{gannon24}, who have applied this estimator to Virgo cluster dwarfs choosing $\gamma_{2D}=1.25$. Assuming spherical symmetry, we deprojected to a 3D density slope according to \cite{alabi16} to $\gamma = \gamma_{\rm 2D} + 1=2.25$.
Decreasing $\gamma$ decreases the inferred dynamical mass overall, while increasing $\gamma$ has the opposite effect.

For $\beta$, we assumed isotropic GC orbits, i.e. $\beta=0$.
$\beta$ can range from $-3 \leq$ $\beta$ $\leq$ 1, where $-3$ describes strongly tangential orbits and 1 describes fully radial orbits. Decreasing $\beta$ increases the dynamical mass, increasing $\beta$ decreases the dynamical mass. \cite{liang24}, when investigating dynamical friction in UDG1, found that in the scenario of a cored halo, orbits can become slightly more radial on average if their orbits are in the vicinity of the core radius. Otherwise, however, there is no expectation of dynamical friction influencing the shape of GC orbits \citep{vandenbosch99}.

For $\alpha$, the chosen parameter corresponds to assumptions about the underlying dark matter halo profile. The inferred dynamical mass increases with $\alpha$. Choosing $\alpha=-1$ corresponds to the assumption that the underlying gravitational potential falls off with a slope of $-1$, i.e. a cuspy NFW profile. Choosing $\alpha=-2$ corresponds to a homogenous sphere generating a harmonic potential \citep{watkins10}, i.e., the core region of a cored DM profile.

We consider the dynamical mass with a cusp ($\alpha = -1$) and a core ($\alpha = -1.9$) as upper and lower limits for the expected range of masses. It lies between the assumption of no core (i.e. a cusp) and a core that stretches out at least to the outermost tracer at 1.63 $R_{\rm eff}$. We used $\beta=0$ and $\gamma=2.25$ for both cusp and core as explained above.

With these parameter choices, 10000 random realisations of the GCs' individual velocities were drawn based on their uncertainties and the dynamical mass was calculated after each draw. The median of the 10000 realisations is reported as the dynamical mass, and the 16th and 84th percentile respectively as the lower and upper uncertainty on the mass.
This way we determined $M_{\rm cusp}$(r < 3.1 kpc)=1.98$^{+0.32}_{-0.28}\times$ 10$^{9}$ M$_{\odot}$ for a cuspy halo and $M_{\rm core}(r < 3.1 \rm kpc)=3.66^{+0.74}_{-0.65}\times$ 10$^{8}$ M$_{\odot}$ for a cored halo. Both values are the mass within the radius of the outermost GC.
The estimates calculated under the assumptions of a core and a cusp represent the upper and lower end of a range of expected dynamical masses, which is shown as a capped blue line in Figure \ref{fig:mdyn_total}.

Additionally, we calculated the dynamical mass enclosed within the deprojected effective radius according to \cite{wolf10}. We note that this estimator inherently makes the assumption of a flat velocity dispersion profile. As shown in Section \ref{sec:results_sigma}, this is consistent with the GC data, however, it might not be the case for the stellar velocity dispersion in UDG1, as is discussed in Section \ref{sec:sigma_discussion}.
With their equation 2,
\begin{equation}\label{eq:m_wolf}
    M(< r_{\rm eff}) \simeq 930 \Biggl( \frac{\sigma^2_{\rm GC}}{\text{km}^2\text{s}^{-2}} \Biggr) \Biggl( \frac{R_{\rm eff}}{\text{pc}} \Biggr) \text{M}_{\odot}, 
\end{equation}
$\sigma_{\rm GC}=29.8^{+6.4}_{-4.9}$ km s$^{-1}$ and assuming the stellar $R_{\rm eff}=1.9$ kpc, we calculated $M_{\rm dyn}$ (r $\leq$ 2.5 kpc) = 2.09$^{+1.00}_{-0.64}\times$ 10$^{9}$ M$_{\odot}$ within the deprojected half-light radius, $r_{\rm eff} \simeq$ 2.5 kpc.
Using the stellar velocity dispersion from \cite{forbes21}, $\sigma_{\ast} = 17 \pm 2$ km s$^{-1}$, we calculated $M_{\rm dyn,\ast} (\rm r < 2.5 \rm kpc)=6.81^{+1.71}_{-1.52}$ $\times$ 10$^{8}$ M$_{\odot}$. This is lower than the same estimate based on the GC system velocity dispersion due to the direct dependence of the dynamical mass on $\sigma^2$.

We then created DM halo profile models following the process outlined in \cite{forbes24} for a cuspy NFW \citep{nfw96} and a cored \citep{read16, read17} case. We note \cite{forbes24} assumed the dimensionless Hubble constant $h=1$ when calculating the concentration parameters for the DM halo fits in their work. Here we used $h=0.7$ consistent with $H_0=70$ km s$^{-1}$ Mpc$^{-1}$, although this causes negligible differences for profiles with the same halo mass.

To summarise the method briefly, we used the recipe of \cite{dicintio14} to derive a NFW DM halo profile depending only on the total halo mass. For the halo concentration, $c_{200}$, we followed \cite{dutton14}:
\begin{equation}\label{eq:c_dutton}
    \log_{10}(c_{200})=0.905 - 0.101 \times \log_{10}(M_{200}/10^{12}h^{-1}M_{\odot})
\end{equation}
We used
\begin{equation}
    g(c)=\ln(1+c) + \Big(\frac{c}{1 + c}\Big)
\end{equation}
in the integrated, mass dependent density profile, leading to the cuspy profile
\begin{equation}
    M(<r)=M_{200}\,g\Big(\frac{r}{r_s}\Big)\frac{1}{g(c_{200})}.
\end{equation}

For the cored profile, we adjusted the halo profile as in \cite{read16}. This adjustment introduces the parameter $n = r_{\rm c}$/$R_{\rm eff}$, which determines the size of the DM core and ranges from 0 to 2.75. $n=2.75$ corresponds to a fully formed core of maximal size \citep{read17} (`full-size core'). For our cored profile we assumed $n = 2.75$, so that our values for a cuspy and a cored profile represent the masses at respective ends of the possible range of core sizes.

The cuspy halo profile is within the joint uncertainties of the estimate from \cite{wolf10} using the GC velocity dispersion, but higher than the estimate using the stellar velocity dispersion. The cored halo profile assumes the largest possible core size and lies marginally below the estimator from \cite{wolf10} when using the GC velocity dispersion, but marginally higher than the \cite{wolf10} estimator when using the stellar velocity dispersion.

The area between the cuspy and the cored profile in Figure \ref{fig:mdyn_total} can be filled without changing the total halo mass if the core size decreases.
For core sizes of $r_{\rm c} \lesssim 2.6 R_{\rm eff}$, the cored profile is consistent with the $\sigma_{\rm GC}$ based dynamical mass estimate from \cite{wolf10}.
For $r_{\rm c} \gtrsim 2 R_{\rm eff}$, the cored profile remains within the uncertainties of the upper end of the dynamical mass range calculated with \cite{watkins10}.
Similarly, changing the total halo mass also influences the agreement between dynamical mass estimates, namely a cuspy halo with reduced halo mass would similarly fit the estimate with \cite{wolf10} based on the stellar velocity dispersion and the estimate with \cite{watkins10} based on the GC velocities.

The exact profile shape is also dependent on the halo concentration. At the same radius, lowering the concentration flattens the profile and decreases the dynamical mass compared to haloes with the same total halo mass and higher concentrations. Similarly, haloes with high concentrations and low total mass can, at the same radius, have similar dynamical masses to haloes with low concentrations and high total masses. We did not further explore this degeneracy here, but note that additional scatter in the DM profiles is expected for varying concentrations, further limiting the ability to analyse the preferred DM halo profile.

For GC-rich UDGs, a cored halo is, however, a prediction from \cite{forbes24}. They found, based on $\sigma_{\ast}$, a cored halo with the halo mass $M_{200} = 4.89~ \times 10^{10}$ M$_{\odot}$ (cusp with $M_{200} = 0.42~ \times 10^{10}$ M$_{\odot}$) for UDG1. This is similar to what common stellar mass--halo mass relationships \citep{moster18, behroozi19, danieli23} suggest when extrapolated down to the stellar mass of UDG1. In order for this total mass to fit the dynamical mass estimates based on the GC data, the concentration would have to be much higher than expected for a dwarf galaxy.
Given this degeneracy, it is not possible to constrain the halo profile, but Figure \ref{fig:mdyn_total} does show that an overly massive halo is a possibility for UDG1. The rich GC system and the higher dynamical mass inferred from it do, however, fit the picture of a failed galaxy in \cite{forbes24} and a cored profile would therefore be expected in UDG1 as well.

\section{Discussion}\label{sec:discussion}
\subsection{Globular Cluster Numbers}\label{sec:ngc_discussion}
We spectroscopically confirmed a GC system containing at least 20 GCs. All confirmed GCs fulfill the GC candidate criteria from D22, who estimated a total GC system of $54 \pm 9$ GCs. Amongst the sources we selected based on the D22 criteria for which we obtained a velocity, we did not find any foreground stars or background galaxies. Notably, the D22 criteria for the GC candidates already required $\rm FWHM>2.1$ pix for GCs with magnitudes $m_{\rm F606W}<25$ mag. This means that sources brighter than 25 mag ($\sim 40$ GC candidates) must be partially resolved, i.e., they cannot be foreground stars.
In Section \ref{sec:resinterlopers}, we discussed the possibility of GC interlopers from nearby giant ellipticals in the group, NGC 5846, NGC 5813 and NGC 5838 and found contamination by intra-group GCs to be very unlikely.

\begin{figure}
    \centering
    \includegraphics[width=\columnwidth]{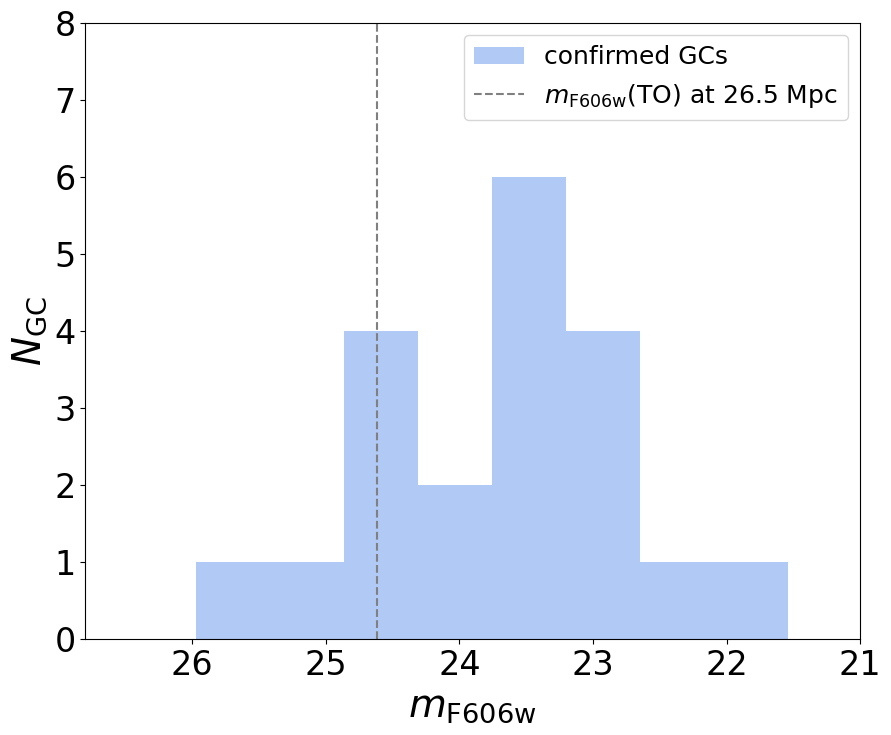}
    \caption{The globular cluster luminosity function (GCLF) for 20 confirmed GCs in apparent magnitudes. Marked with a dashed line is the expected turnover magnitude $M_{V}(\rm TO)=-7.5$ mag of the GCLF assuming a distance of $d=26.5$ Mpc. After correcting for missing area coverage in the brighter half, the minimum number of GCs is expected to be $N_{\rm GC} > 35$. Approximately $9$\% of the stellar luminosity is contained in the 20 confirmed GCs.}
    \label{fig:gclf}
\end{figure}
A lower limit to $N_{\rm GC}$ can be provided by making use of the globular cluster luminosity function (GCLF). Remarkably, the GCLF is almost the same for galaxies of different types across different environments \citep{richtler03, miller-lotz07}. It has a Gaussian shape and displays characteristic values for the turnover magnitude, $M_{V}$(TO), and spread, as suggested by \cite{hanes77}.
Since the GCLF is symmetric, the total number of GCs can be estimated from this if the GCs in the brighter half are simply doubled as, for example, in \cite{forbes21}. 

Figure \ref{fig:gclf} shows the GCLF for spectroscopically confirmed GCs from this work. D22's best fitting turnover absolute magnitude, $M_{\rm F606W}(\rm TO)=-7.5$ mag, corresponds to the turnover apparent magnitude $m_{\rm F606W}(\rm TO)=24.6$ mag at their assumed distance. This magnitude is marked in Figure \ref{fig:gclf} as a dashed line.
We use this GCLF turnover apparent magnitude to provide a lower limit on $N_{\rm GC}$, using the symmetry of the Gaussian GCLF. In this work we used D22's candidate list as it is from the deepest available imaging.
Before computing the expected minimum number of GCs, we corrected for missing area in the bright half of the GCLF. Approximately 40 per cent of UDG1's area within 2 $R_{\rm eff}$ is covered by our pointings. Within this covered area, 15 of 22 GC candidates brighter than $m_{F606W}(\rm TO)=24.6$ mag are spectroscopically confirmed. Assuming the same rate of confirmation in the regions we did not probe, 2.7 additional GCs with $m_{\rm F606W}<m_{F606W}$(TO) are expected within 2$R_{\rm eff}$.
Therefore, we expect at least $15 + 2.7=17.7$ GCs in the GC system with magnitudes brighter than the turnover, and hence for the whole GC system $N_{\rm GC} > 35$. We note that there was no magnitude correction applied as different overlapping pointings result in multiple different limiting magnitudes across the covered area (see Figure \ref{fig:pointings}). The lower limit of $N_{\rm GC} > 35$ is compatible with the estimates by \cite{marleau24} ($N_{\rm GC}=38\pm7$), \cite{forbes21} ($N_{\rm GC}\sim$45) and D22 ($N_{\rm GC}=54\pm9$), but not with the $26\pm6$ GCs estimated by \cite{mueller21}. Given the absence of contaminants in the D22 candidates, $N_{\rm GC} = 54 \pm 9$ is the preferred estimate, independent of the assumed distance.

UDG1 is therefore considered GC-rich by many common definitions for UDGs (e.g. $N_{\rm GC}$ > 20 in \cite{gannon22, buzzo24}). High numbers of GCs ($N_{\rm GC} \gtrsim 20$) are expected for failed galaxies \citep{buzzo22, buzzo24} along with old stellar populations. M20 and \cite{ferre-mateu23} found old stellar populations for UDG1, and it is confirmed in this work to be GC-rich. M20 also measured ages and metallicities for individual GCs and find them agreeing with each other and the stellar populations of the galaxy, indicating formation at the same time as the stellar body. Using just the 20 spectroscopically confirmed GCs, we find the GC system to contain at least $\sim9$\% of UDG1's stellar luminosity. Assuming an average GC dissipation rate, this implies that the galaxy formed the majority of its stars in dense star clusters (D22). A GC fraction this high is expected for a failed galaxy \citep{forbes2024_mgc_citation}, as is the massive DM halo \citep[see also Section \ref{sec:result_mass}]{forbes24}, overall making UDG1 fit the picture of a failed galaxy.

\subsection{Velocity Dispersion}\label{sec:sigma_discussion}
UDG1's stellar ($\sigma_{\ast} = 17 \pm 2$ km s$^{-1}$, \cite{forbes21}) and GC system velocity dispersion ($\sigma_{\rm GC} = 29.8^{+6.4}_{-4.9}$ km s$^{-1}$) do not agree with each other. In all other UDGs for which both values are known, however, they do agree. This would therefore be expected to be the case for UDG1 as well.
A possible reason for this could be an increase in the stellar velocity dispersion with radius. It has only been measured from within $\sim 0.5 R_{\rm eff}$, whereas the GCs in this work trace the potential out to a radius of $\sim 1.6 R_{\rm eff}$.
In the UDG DF 44 in the Coma cluster, the stellar velocity dispersion has been measured in radial bins and shown to have a rising stellar velocity dispersion profile \citep{dokkum19_df44}. DF 44, like UDG1, is also GC rich and has a massive DM halo. Although due to a lack of offset sky exposures it is not possible to measure the stellar velocity dispersion out to further radii with the avialable data, given the similarities to DF 44, it is feasible that UDG1 could also display a rising stellar velocity dispersion profile. This could lead to the expected agreement between $\sigma_{\ast}$ and $\sigma_{\rm GC}$.

Using the stellar and GC system velocity dispersion, we evaluate the DM halo models of \cite{liang24}. They require $\sigma_{\rm GC} < \sigma_{\ast}$ for all their models. They used an MCMC algorithm to fit a cuspy NFW profile and a cored Burkert halo \citep{burkert95} to the result of their semi-analytic modelling of the GC system's evolution, i.e. to the observed spatial present-day GC distribution. They reported the mode and the median value of the velocity dispersion from their MCMC fits of both halo models.
Their cuspy model requires a GC system velocity dispersion of approximately 11 km s$^{-1}$ (mode and median), which does not agree with our measured value of $29.8^{+6.4}_{-4.9}$ km s$^{-1}$ and can be ruled out based on that disagreement.
Their cored model has higher values for the median GC system velocity dispersion ($\sim24$ km s$^{-1}$ at r$\sim2$ kpc). The median value of the stellar velocity dispersion ($\sim25$ km s$^{-1}$) is in this case also consistent with our $\sigma_{\rm GC}$, but not with the stellar velocity dispersion from \cite{forbes21}. Their mode GC system velocity dispersion, however, is still not consistent with our $\sigma_{\rm GC}$ at all.
Unless $\sigma_{\ast}$ almost doubles with increasing radius, \cite{liang24}'s cored model is also incompatible with the observed velocity dispersions as it requires the stellar velocity dispersion to be higher than the GC system velocity dispersion.

\subsection{Dynamical Friction}\label{sec:df_discussion}
One possible physical reason for the increase in velocity dispersion with the addition of fainter (less massive) GCs that we described in Section \ref{sec:results_sigma} is dynamical friction (DF). DF describes the gravitational drag exerted on an object by the stellar body and the DM halo it moves through. The effect of dynamical friction is roughly proportional to the mass of the affected object and the mass ratio to the halo potential \citep{chandrasekhar43}.
In the context of GC systems, DF leads to the following predictions:
\begin{enumerate}
    \item Mass segregation: the influence of DF is expected to make the GCs migrate inwards. Due to the proportionality to the mass, more massive GCs migrate on a shorter timescale than less massive ones \citep{lotz01}. This leads to mass segregation of the GCs, observable as a segregation where brighter GCs are on average closer to their host galaxy's centre than fainter GCs.
    \item Formation of a nucleus: the expected timescale of GCs migrating inwards is dependent on the assumed halo profile, but especially in cuspy haloes is generally short compared to the life time of most galaxies \citep{lotz01}. Eventually, under the influence of DF, the GCs are expected to sink into the centre and merge into a nuclear star cluster \citep{tremaine76, lotz01, Oh00, sanchez-salchedo06}. The absence of a nucleus, however, does not imply that DF has no significant effect in any given GC system. The inward migration can be stalled at the core radius in a cored DM halo \citep{goerdt06}, or prevented in a cuspy halo if the GCs initially form at very large radii \citep{bar22}.
    \item GC system/stellar velocity dispersion: For UDG1, \cite{liang24} presented a semi-analytical model of GC evolution under the effect of DF in a composite host potential consisting of baryonic and DM contributions.
    They investigated the velocity dispersions of GCs and the stellar body of UDG1 for a cuspy NFW \citep{nfw96} and for a cored Burkert \citep{burkert95} DM halo model. For both halo models, they found $\sigma_{\rm GC}$ lower than or similar to $\sigma_{\ast}$ (see their figure 9). Their exact predicted values for $\sigma_{\rm GC}$ and $\sigma_{\ast}$ do not match the values from M20 and \cite{forbes21}, but qualitatively they made a general prediction of $\sigma_{\rm GC}$ < $\sigma_{\ast}$. 
    \item Velocity dispersion with radius: `Perfect' mass segregation would lead to an increasing radial velocity dispersion profile in three dimensions, in agreement with the increase with magnitude. However, in projection $\sigma_{\rm GC}$ is expected to remain flat or even decrease with increasing radii \citep{bilek19, liang24}.
\end{enumerate}

\begin{table}
    \centering
    \begin{tabular}{l|c}
        DF expectation & Observed in UDG1 \\
        \hline
        \hline
        Mass segregation & $\checkmark$ \\
        Formation of a nucleus & $\text{\sffamily X}$ \\
        $\sigma_{\rm GC} < \sigma_{\ast}$ & $\text{\sffamily X}$ \\
        $\sigma_{\rm GC}$ increasing with magnitude & $\checkmark$ \\
        $\sigma_{\rm GC}$ decreasing with radius & $\text{\sffamily X}$ \\
        \hline
    \end{tabular}
    \caption{A summary of the expected properties from dynamical friction compared to the observations of UDG1.}
    \label{tab:df_predicitons}
\end{table}

These predictions can be addressed by our results:
\begin{enumerate}
    \item Mass segregation: The trend of mass segregation originally found by \cite{bar22} is shown in the top panel of Figure \ref{fig:sig_by_mag}, showing the average radii of GCs with magnitudes corresponding to the bins in which the velocity dispersion was calculated, both for the spectroscopically confirmed GCs (blue circles) and GC candidates from D22 (purple Ys, restricted to GCs within 2 $R_{\rm eff}$). The average radius of the imaging candidates increases with fainter magnitudes (decreasing mass), whereas the increase for the confirmed GCs flattens, likely due to a strong observational bias. Fainter GCs require longer exposure times and could therefore only be confirmed in the central region of UDG1 where multiple pointings overlap (see Figure \ref{fig:pointings}, Table \ref{tab:observations_overview}). The increase in the average radius of the GC candidates, however, matches the predictions of DF: fainter and brighter GCs could have started out with a similar radial distribution with DF causing the brighter (massive) GCs to migrate inwards on a shorter time scale compared to the fainter ones. This scenario is consistent with the findings of both \cite{bar22} and \cite{liang24}. The observed radial mass segregation could also be reinforced by an initial mass--radius trend existing in the GC system such as exists for more massive galaxies \citep[see e.g. figure 7 in][]{baumgardt19}.
    \item Formation of a nucleus: UDG1's brightest GC (GC 1 in Table \ref{tab:gc_list}) could be a nucleus based on its location close to the galaxy's centre, its brightness and its previously measured velocity \citep{mueller20, forbes21}. We note that the velocity we find for GC 1 ($v$=2137.8 $\pm$ 4.5 km s$^{-1}$) is lower than the galaxy velocity of 2156.2 $\pm$ 2 km s$^{-1}$ \citep[corrected by measured offset, see Appendix \ref{app:bh3m_vs_bh3l}]{forbes21}, even after taking into consideration possible differences in the wavelength calibration (see Appendix \ref{app:bh3m_vs_bh3l}). This makes it unlikely to be a nucleus, although the possibility cannot be clearly ruled out. As explained above, the absence of a nucleus does, however, not contradict the predictions of DF.
    \item GC system/stellar velocity dispersion: In contrast to \cite{liang24}'s prediction based on DF, we find $\sigma_{\rm GC} > \sigma_{\ast}$ for the whole GC system.
    To better compare $\sigma_{\ast}$ and $\sigma_{\rm GC}$ in UDG1, we determined the velocity dispersion of the GCs within the same area as the stellar light, which was measured from within $\sim 0.5 R_{\rm eff}$, and found $\sigma_{\rm GC, F21 area}=25.9^{+10.5}_{-6.6}$ km s$^{-1}$ (listed also in Table \ref{tab:velocity_dispersions}). This is within the joint uncertainties of $\sigma_{\ast}=17 \pm 2$ km s$^{-1}$. However, it leans noticeably higher. As already mentioned, this is unusual for UDGs, for which these two values were found to be generally the same \citep{forbes21}.
    The prediction of $\sigma_{\rm GC} < \sigma_{\ast}$  in \cite{liang24} is made specifically for UDG1. A lack of agreement with the measured values could imply either that their model is not applicable as it is to UDG1, that other effects (e.g. the initial distribution of GC magnitudes) outweigh DF, or that the stellar velocity dispersion increases with radius, as already discussed in Section \ref{sec:sigma_discussion}.  
    
    The GC velocity dispersion $\sigma_{\rm GC}$ is dependent on the magnitude of the GCs for UDG1, as shown in the bottom panel of Figure \ref{fig:sig_by_mag}. In the bottom panel, it is shown that some bins containing brighter GCs do have $\sigma_{\rm GC} < \sigma_{\ast}$, which suggests that the selection of a brighter subsample can lead to agreement with this prediction. The GC velocity dispersion is, however, mostly consistent with the stellar velocity dispersion for the brightest 14 GCs down to an apparent magnitude of $\sim$23.8 mag and increases only when fainter GCs are added. An increase of $\sigma_{\rm GC}$ with fainter GCs is expected, as DF is expected to decrease the dispersion for the more strongly affected, brighter GCs.
    \item Velocity dispersion with radius: The moving window profile of velocity dispersion with radius (see Figure \ref{fig:sig_by_r}) is flat, although not inconsistent with a decrease at larger radii.
    With the sensitivity of $\sigma_{\rm GC}$ to outliers (see Section \ref{sec:results_sigma}) and some of the faintest GCs having very small projected radii, it is not possible to track the true change of the velocity dispersion with the three dimensional radius, r, via the projected radius, R.
\end{enumerate}

Table \ref{tab:df_predicitons} shows a summary of the DF predictions and corresponding observations in UDG1. The absence of a nucleus can be caused by a cored DM halo, as is favoured for UDG1 (see Section \ref{sec:result_mass}). We cannot sufficiently test the radial velocity dispersion profile, but the observed increase of the velocity dispersion with increasing magnitude is expected for DF. Overall, combined with the mass segregation trend observed in \cite{bar22}, we find evidence that dynamical friction is relevant to UDG1's GC system. In contrast to \cite{liang24}'s predictions, however, we find $\sigma_{\rm GC} > \sigma_{\ast}$, which requires investigation in future work.

\section{Summary}\label{sec:summary}
In this work, we studied the globular cluster system of NGC5846\_UDG1 with spectroscopic data from KCWI on the Keck telescope. We confirmed 19 GCs as members of the galaxy. Combined with the GC sample from \cite{mueller20}, a total of 20 GCs are now spectroscopically confirmed to be members of the galaxy, with no contaminants found in D22's imaging candidates. We found the following:
\begin{itemize}
    \item Approximately 9\% of UDG1's stellar light is contained in the 20 confirmed GCs.
    \item After correcting for missing area of coverage, we found the lower limit on the number of GCs for UDG1 to be $N_{\rm GC} \geq 35$. This minimum is based on the GCLF and on the assumption of $d=26.5$ Mpc and the criteria for imaging candidates in \cite{danieli22}, who estimated the total number to be $N_{\rm GC}=54\pm 9$ GCs from deep \textit{HST} imaging.
    None of the confirmed GCs are expected to be intra-group GCs, specifically interlopers from one of the dominant, giant ellipticals in the group, NGC 5813 or NGC 5846 or from the close-in-projection giant elliptical NGC 5838.
    \item The GC system velocity dispersion for the 20 confirmed GCs is $\sigma_{\rm GC}=29.8^{+6.4}_{-4.9}$ km s$^{-1}$, with the mean velocity of $\bar{v}_{\rm GC}=2153.9^{+7.1}_{-7.0}$ km s$^{-1}$. Within the sample, the velocity dispersion increases with increasing GC magnitudes and remains flat to $\sim1R_{\rm eff}$.
    \item Our findings on the GC velocity dispersion, combined with previous results from \cite{bar22}, are mostly consistent with the expectation from dynamical friction. Namely there is mass segregation in the GC system, although there is no nucleus. Bright GCs have a lower velocity dispersion than fainter ones. For the brightest GCs, the velocity dispersion is also lower than the stellar velocity dispersion, however, this does not hold for the whole GC system.
    \item We derived dynamical mass estimates from the GC velocity dispersion, finding $M_{\rm dyn}=2.09^{+1.00}_{-0.64}\times 10^{9}$M$_{\odot}$ within the de-projected half-light radius $r_{\rm eff} \simeq 2.5$ kpc.
    \item The total halo mass suggested by the $N_{\rm GC}-M_{200}$ relationship using the 54 GC candidates from D22, $M_{200}=2.7^{+2.7}_{-1.4}~ \times 10^{11}$ M$_{\odot}$, is higher than masses suggested by the SMHM relation. Both a cuspy and a cored halo profile with this mass are, however, consistent with the dynamical mass we measured from the GC velocity dispersion.
    \item UDG1, with an overly massive, likely cored, halo, a rich GC system, and a high GC luminosity fraction, fits the picture of a failed galaxy.
\end{itemize}

\section*{Acknowledgements}
We would like to thank the anonymous referee for their careful read of the paper and their suggestions, which significantly improved the manuscript.
We would like to thank Sergio Guerra Arenciba for sharing his work comparing a preliminary selection of GCs in UDG1 to imaging data. We also thank Warrick Couch and Ned Taylor for helpful discussions during this project.
This research was supported by the Australian
Research Council Centre of Excellence for All Sky Astrophysics in 3 Dimensions (ASTRO 3D), through project number CE170100013.
DF thanks the ARC for support via DP220101863 and DP200102574.
AJR was supported by National Science Foundation Grant AST-2308390. This work was supported by a NASA Keck PI Data Award, administered by the NASA Exoplanet Science Institute.
This research made use of Montage. It is funded by the National Science Foundation under Grant Number ACI-1440620, and was previously funded by the National Aeronautics and Space Administration's Earth Science Technology Office, Computation Technologies Project, under Cooperative Agreement Number NCC5-626 between NASA and the California Institute of Technology.
This research has made use of the Keck Observatory Archive (KOA), which is operated by the W. M. Keck Observatory and the NASA Exoplanet Science Institute (NExScI), under contract with the National Aeronautics and Space Administration. 
Some of the data presented herein were obtained at Keck Observatory, which is a private 501(c)3 non-profit organization operated as a scientific partnership among the California Institute of Technology, the University of California, and the National Aeronautics and Space Administration. The Observatory was made possible by the generous financial support of the W. M. Keck Foundation. The authors wish to recognize and acknowledge the very significant cultural role and reverence that the summit of Maunakea has always had within the Native Hawaiian community. We are most fortunate to have the opportunity to conduct observations from this mountain.

\section*{Data Availability}
KCWI data is available 18 months after observations via the Keck Observatory Archive (KOA): https://www2.keck.hawaii.edu/koa/public/koa.php.

\section*{Software}
astropy \citep{astropy13, astropy18}, ppxf \citep{cappellari04, cappellari2022}, emcee \citep{foreman-mackey13}, montagepy



\bibliographystyle{mnras}
\bibliography{example} 




\appendix

\section{KCWI Medium and Large 
Slicer}\label{app:bh3m_vs_bh3l}
\begin{figure}
    \centering
    \includegraphics[width=\columnwidth]{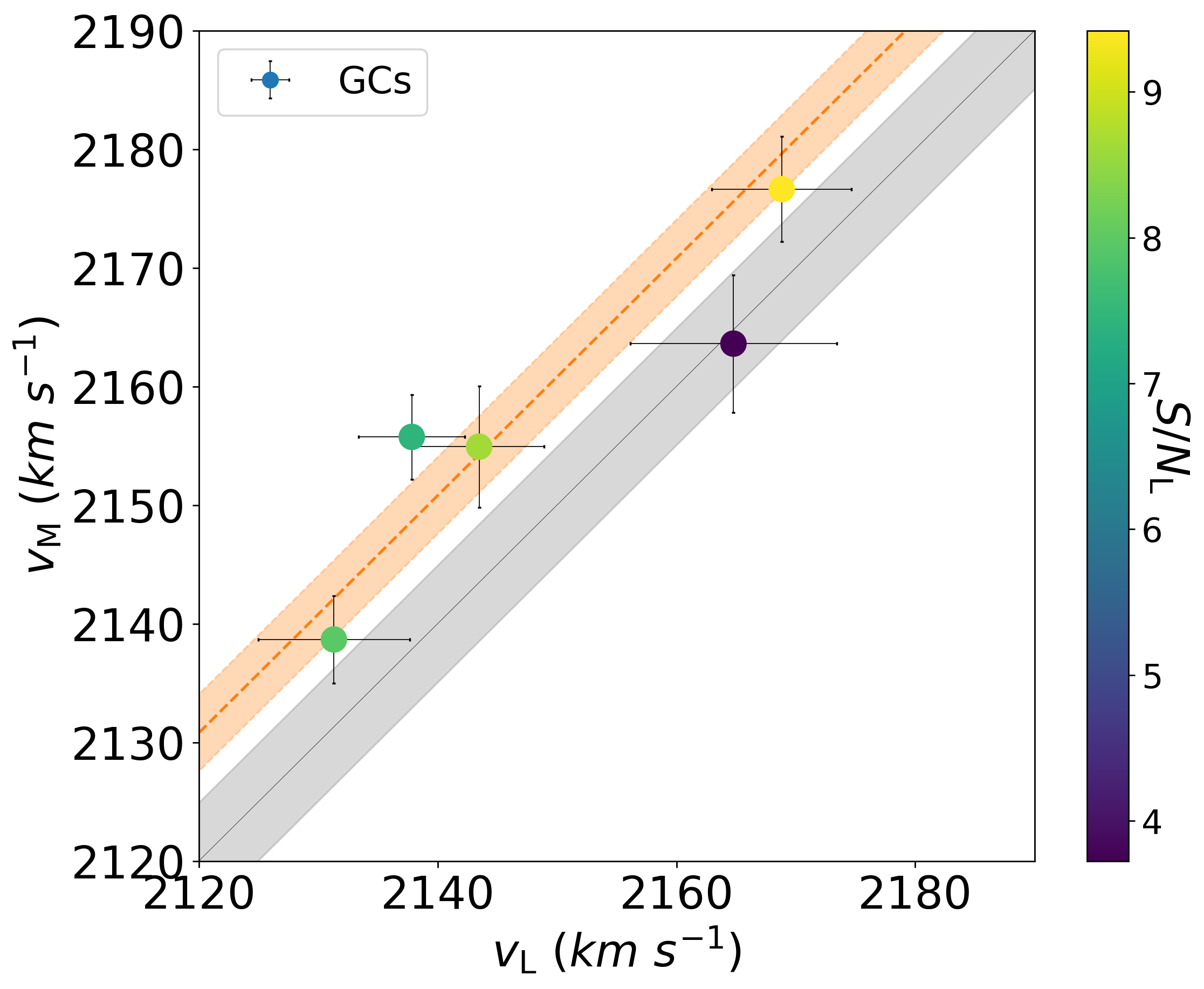}
    \caption{Recessional velocities of 5 GCs in common between the Medium and Large slicers from KCWI, colour coded by the S/N ratio measured from the Large slicer. The black line shows a  one-to-one relationship with S/N weighted 1$\sigma$ scatter ($\pm$ 4.9 km s$^{-1}$) shaded in grey around it. The orange dashed line shows the best fit for a potential systematic offset between the slicers (10.8 $\pm$ 3.3 km s$^{-1}$). The offset was applied to all velocities measured only in the Medium slicer throughout this work.}
    \label{fig:bh3m_vs_bh3l}
\end{figure}
Out of the 19 GCs, 5 yielded spectra fulfilling all of the criteria described in Section \ref{sec:data_analysis} from both the Medium and the Large slicer. The values from both slicers are listed in Table \ref{tab:bh3m_bh3l}; Figure \ref{fig:bh3m_vs_bh3l} shows the velocities measured in both slicers for these GCs.
\begin{table}
    \centering
    \begin{tabular}{c|c|c}
    \hline
        GC & $v_{\rm M}$ & $v_{\rm L}$\\
          & [km s$^{-1}$] & [km s$^{-1}$]\\
        \hline
        \hline
        1 & 2155.7$\pm$3.8 & 2137.8$\pm$1.7\\
        2 & 2154.9$\pm$5.2 & 2143.5$\pm$2.1\\
        4 & 2176.6$\pm$4.5 & 2168.8$\pm$2.9\\
        7 & 2138.7$\pm$3.7 & 2131.3$\pm$6.4\\
        8 & 2163.6$\pm$5.9 & 2164.8$\pm$8.7\\
        \hline
    \end{tabular}
    \caption{Recessional velocities and corresponding uncertainties from the Medium (M) and Large (L) slicer respectively for those GCs where values were obtained from both slicers.}
    \label{tab:bh3m_bh3l}
\end{table}

To test for a potential systematic offset, we performed a $\chi^2$-minimisation to fit a straight line with a fixed slope of unity. The best fit value of the offset is $10.8 \pm 3.3$ km s$^{-1}$, shown as an orange dashed line in Figure \ref{fig:bh3m_vs_bh3l}. We tested the significance of the offset with different statistical tests:
\begin{itemize}
    \item A $\chi^2$-test: The $\chi^2$ value of the offset is 3.27. We tested the likelihood for a smaller $\chi^2$ and found $p=0.35$, indicating that the offset is not significant.
    \item A Kolmogorov--Smirnov (KS) test: We performed a two-sample KS test of the offset between the two distributions (velocities measured with the Medium and Large slicer, respectively), and found $p=0.8$, indicating that the offset is not significant.
    \item Consistency with a Gaussian distribution: Figure \ref{fig:bh3m_vs_bh3l} shows a hypothetical 1-to-1 relationship between the two velocity distributions (black line). For this we measured a 1$\sigma$ scatter of 4.9 km s$^{-1}$ (shaded grey area). We tested whether the difference between the Medium and Large velocities is significantly different from a Gaussian distribution with $\sigma$=4.9 km s$^{-1}$ with a one-sided KS test. For this we found $p=0.0028$, indicating that the velocities are not consistent with being drawn from a single Gaussian distribution, i.e. the offset is significant.
    \item Significance of bimodality according to \cite{ashman94}: We tested for the significance of the separation between the Medium and the Large velocities and found $D=3.11$. Values of $D>2$ indicate a significant separation, therefore indicating that the Medium and Large velocities are not consistent with a single distribution.
\end{itemize}
Since the Large slicer yielded a recessional velocity for all but two GCs, the velocities measured from the Large slicer are listed as the final velocity in Table \ref{tab:gc_list} whenever possible. GC 14 and GC 18 in Table \ref{tab:gc_list} only yielded a reliable fit from the Medium slicer, which was corrected with the offset of $-10.8$ km s$^{-1}$ and listed as the final velocity in Table \ref{tab:gc_list}.

With two tests showing the measured offset as insignificant and two tests showing significance, we treated the offset as significant, but report the results without the offset applied in Appendix \ref{app:no_offset}. There are small differences to the velocity dispersion of the GC system, but the overarching conclusions do not change depending on whether an offset is applied or not.

\cite{forbes21} also used the KCWI BH3 grating with the Medium slicer to measure the velocities for two GCs which are measured in this work with the same setup. They reduced the data using the KCWI IDL data reduction pipeline, whereas we use the Python pipeline for the reduction of the same data. For their `GC 9' and `GC 10', corresponding to our `GC 1' and `GC 2', respectively, we get a velocity $\sim$10 km s$^{-1}$ lower than their value. With just two sources to compare, we do not conduct the same analysis for a systematic offset, but do caution that this might create lower mean GC velocities in comparison to the galaxy recessional velocity measured from the stellar light in \cite{forbes21}.

\section{Comparison to Literature}\label{app:kcwi_vs_muse}
\begin{figure}
    \centering
    \includegraphics[width=\columnwidth]{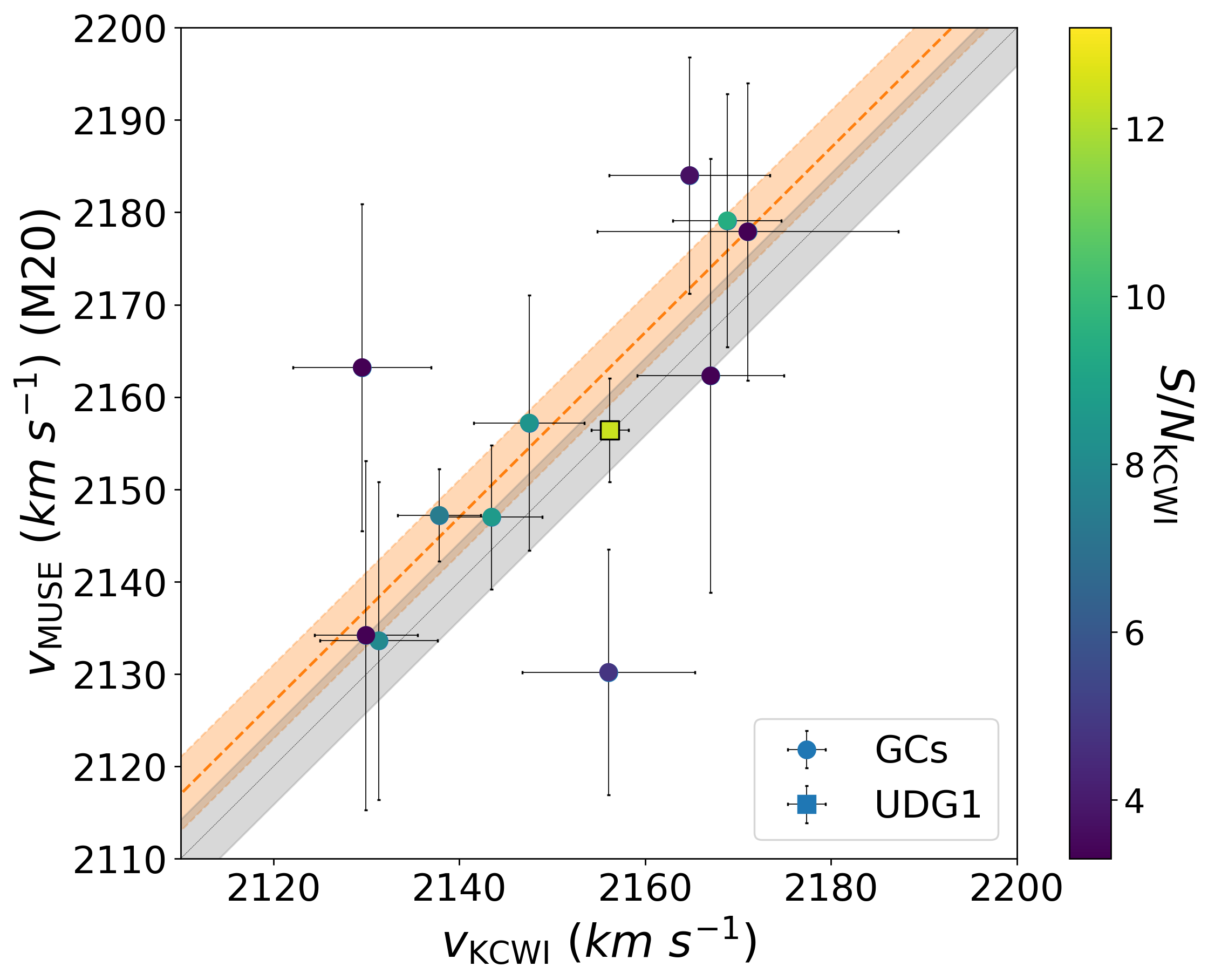}
    \caption{Recessional velocities from M20 (MUSE) and from this work (KCWI), colour coded by KCWI S/N ratio. The round symbols show the recessional velocities of the 11 GCs contained in both samples. The square symbol shows the recessional velocity for UDG1 itself from M20 and \protect\cite{forbes21} (not included in the fit for the offset). The black line shows a one-to-one relation with the S/N weighted 1$\sigma$ scatter ($\pm$ 4.2 km s$^{-1}$) shaded in grey around it. The orange dashed line shows the best fit for a potential systematic offset between M20 and this work (7.0 $\pm$ 4.1 km s$^{-1}$). The offset was applied to `GC 2 (M20)' throughout this work.}
    \label{fig:kcwi_vs_mueller}
\end{figure}
Eleven GCs had been previously confirmed to be within 100 km s$^{-1}$ of UDG1's recessional velocity by M20 with the MUSE spectrograph on the Very Large Telescope in Chile. We recovered ten of those eleven GCs in this work. The eleventh GC (`GC 2' in M20) is not within the covered area of the KCWI pointings. Figure \ref{fig:kcwi_vs_mueller} shows the recessional velocities for GCs covered in both M20 and this work, measured with MUSE and KCWI, respectively. The velocity of UDG1 itself from M20 (MUSE) and \cite{forbes21} (KCWI, Medium slicer) is included in Figure \ref{fig:kcwi_vs_mueller} as well, but not used in the analysis of the offset. 

M20 also listed a joint velocity for two further GC candidates based on their two spectra stacked together, where the S/N of the individual spectra did not allow for recovering a recessional velocity. Of these two candidates we were able to isolate and confirm one (`cand 2' in M20) as a GC, which is listed as GC 8 in Table \ref{tab:gc_list}. The second candidate was affected by a severe noise spike at the redshift at which we would expect the H$_{\beta}$ line, and we were not able to ensure that the measured velocity was not dominated by this, hence we do not report it as a confirmed GC.

We tested for a systematic offset between M20 and this work and its statistical significance the same way as was done for the two KCWI slicers in Appendix \ref{app:bh3m_vs_bh3l}. For the offset we found $-7 \pm 4.1$ km s$^{-1}$. 
\begin{itemize}
    \item A $\chi^2$-test: The $\chi^2$ value of the offset is 7.3. We tested the likelihood for a smaller $\chi^2$ and found $p=0.61$, indicating that the offset is not significant.
    \item A Kolmogorov-Smirnov (KS) test: We performed a two-sample KS test testing whether the offset between the two distributions (velocities measured with MUSE and KCWI, respectively), and find $p=0.8$, indicating that the offset is not significant.
    \item Consistency with a Gaussian distribution: Figure \ref{fig:bh3m_vs_bh3l} shows a hypothetical 1-to-1 relationship between the two velocity distributions. For this we measured a 1$\sigma$ scatter of 4.2 km s$^{-1}$. We tested whether the difference between the MUSE and the KCWI velocities is significantly different from a Gaussian distribution with $\sigma=4.2$ km s$^{-1}$ with a one-sided KS test. For this we found $p=0.0023$, indicating that the velocities are not consistent with being drawn from a single Gaussian distribution.
    \item Significance of bimodality according to \cite{ashman94}: We tested for the significance of the separation between the MUSE and the KCWI velocities and found $D=2.36$, indicating that the two sets of measurements are not consistent with a single distribution.
\end{itemize}
As with the GCs measured in different slicers, we found the offset between velocities measured with MUSE and measured with KCWI to be insignificant according to the $\chi^2$ test and the two-sample KS test, but significant according to a consistency test with a single Gaussian distribution and the \cite{ashman94} way of determining significance of separation. We therefore also applied the offset to the GC measured only in M20 and include it in our final GC sample with the modified velocity. It is listed as `GC 2 (M20)' in Table \ref{tab:gc_list}

\section{Results without Velocity Offsets}\label{app:no_offset}
The offsets determined in Appendix \ref{app:bh3m_vs_bh3l} and \ref{app:kcwi_vs_muse} were applied to the respective measurements throughout this work, namely two GCs measured only in the Medium slicer and one GC measured only with MUSE in M20. 
However, some of the performed tests showed the offset to not be significant. Therefore, we describe here how the results change if the offset is not applied to those three GCs.

The velocity dispersion for the whole system is $\sigma_{\rm GC} = 27.7^{+6.0}_{-4.6}$ km s$^{-1}$ instead of $\sigma_{\rm GC} = 29.8^{+6.4}_{-4.9}$ km s$^{-1}$. The velocity dispersion profile with magnitude shown in Section \ref{sec:results_sigma} remains rising with increasing magnitude and the profile with increasing radius remains flat.
The dynamical mass calculated based on the non-corrected velocities remains within errors of the current values, although slightly lower. It is within joint uncertainties with the cored DM profile shown in Figure \ref{fig:mdyn_total}.

Qualitatively, the results do not change depending on whether the offset is applied or not, namely
\begin{itemize}
    \item $\sigma_{\rm GC}$ of the whole GC system is higher than $\sigma_{\ast}$ from the stellar light,
    \item the velocity dispersion profile rises with increasing magnitude and is flat with increasing radius,
    \item the dynamical masses with or without the offsets are within errors with each other, however, the dynamical mass without the offsets is also within joint uncertainties of a full-size DM core with the total mass suggested by the $N_{\rm GC}-M_{200}$ relationship.
\end{itemize}
The exact values with and without the offset are listed in Table \ref{tab:offset_comparison}.
\begin{table}
    \centering
    \begin{tabular}{c|c|c}
        \hline
         & Offset & No offset \\
        \hline
        \hline
        $\sigma_{\rm GC}$ [km s$^{-1}$] & 29.8$^{+6.4}_{-4.9}$ & 27.7$^{+6.0}_{-4.6}$\\
        $\bar{v}_{\rm GC}$ [km s$^{-1}$] & 2153.9$^{+7.1}_{-7.0}$ & 2155.0$^{+6.7}_{-6.6}$\\
        $M_{\rm dyn}$ \citep{wolf10} [M$_{\odot}]$& 2.09$^{+1.00}_{-0.64}\times$ 10$^{9}$ & 1.81$^{+0.87}_{-0.56}\times$ 10$^{9}$\\
        \hline
    \end{tabular}
    \caption{The results of the kinematic analysis with and without offsets between different measurements applied as described in Appendix \ref{app:bh3m_vs_bh3l} to \ref{app:no_offset}. From top to bottom we list the whole GC system's velocity dispersion, $\sigma_{\rm GC}$, the system's mean velocity, $\bar{v}_{\rm GC}$, and the dynamical mass calculated with \protect\cite{wolf10}.}
    \label{tab:offset_comparison}
\end{table}


\bsp	
\label{lastpage}
\end{document}